\documentclass[twocolumn,floats,floatfix,amssymb,nofootinbib,prd,superscriptaddress,aps]{revtex4-1}
\usepackage{graphicx,amssymb,amsmath,amsthm,amsfonts,epsfig}
\usepackage[linktocpage,colorlinks=true,citecolor=OliveGreen,linkcolor=Maroon,urlcolor=Maroon]{hyperref}
\usepackage[usenames]{color}
\usepackage{bm}
\usepackage{dcolumn}
\usepackage[utf8]{inputenc}
\usepackage{latexsym}
\usepackage{rotating}
\usepackage{tabularx}
\usepackage{braket}
\usepackage{esvect}
\usepackage{enumerate}
\usepackage{footmisc}
\usepackage{array}
\usepackage{tensor}
\usepackage{mathtools}
\usepackage{url}
\usepackage[dvipsnames]{xcolor}
\usepackage{comment}
\usepackage{multirow}
\usepackage{nicematrix}
\usepackage{graphics}
\usepackage{textcomp}
\usepackage{stmaryrd}
\usepackage[normalem]{ulem}
\usepackage{verbatim}
\usepackage{soul,xcolor}
\usepackage{xspace}
\usepackage{xfrac}
\usepackage{ragged2e}
\setstcolor{red}
\usepackage{esint}
\usepackage{physics}
\usepackage{subcaption}

\def\be{\begin{equation}}
\def\ee{\end{equation}}

\newcommand{\beq}{\begin{eqnarray}}

\newcommand{\eeq}{\end{eqnarray}}

\def\be{\begin{equation}}
\def\ee{\end{equation}}
\def\bea{\begin{eqnarray}}
\def\eea{\end{eqnarray}}
\def\beq{\begin{eqnarray}}
\def\eeq{\end{eqnarray}}

\begin{document}

\title{The relativistic restricted three-body problem:\\
geometry and motion around tidally perturbed black holes}

\author{Takuya Katagiri}
\affiliation{Dipartimento di Fisica, Sapienza Università di Roma, Piazzale Aldo Moro 5, 00185, Roma, Italy}
\affiliation{INFN, Sezione di Roma, Piazzale Aldo Moro 2, 00185, Roma, Italy}

\author{Vitor Cardoso}
\affiliation{Center of Gravity, Niels Bohr Institute, Blegdamsvej 17, 2100 Copenhagen, Denmark}

\affiliation{CENTRA, Departamento de F\'{\i}sica, Instituto Superior T\'ecnico -- IST, Universidade de Lisboa -- UL,
Avenida Rovisco Pais 1, 1049-001 Lisboa, Portugal}

\date{\today}

\begin{abstract}
We investigate the geometry of a tidally deformed, rotating black hole and timelike geodesics in its vicinity. Our framework provides a local picture of the structural evolution of a relativistic restricted three-body problem around a deformed black hole in an adiabatically evolving binary, motivated by various astrophysical settings including disk dynamics and extreme mass-ratio inspirals. As the tidal-field strength is increased, initially regular, bound geodesics undergo four stages:~(i)~weak chaos emerges within the bound motion; (ii)~a subset of trajectories plunges into the black hole;~(iii)~a fraction of the remaining trajectories becomes unbound; and (iv)~no bound trajectories persist. We provide semi-analytic estimates for the critical tidal amplitudes associated with each transition. Our estimates, within the idealized test-particle description, indicate that, within the frequency band of ground-based gravitational-wave detectors, the matter flow around black holes may already be depleted, whereas LISA and (B-)DECIGO could probe the earlier stages. Our results suggest that an object orbiting a tidally deformed massive black hole may remain near resonances in a long term, indicating an accumulated, non-negligible impact on the gravitational-wave phase. Another finding is that tidal perturbations can modulate nonlinear couplings among epicyclic oscillations of geodesics, and could therefore, in principle, affect resonant excitation mechanism potentially relevant to quasi-periodic oscillations in X-ray light curves from accreting black holes. 
\end{abstract}

\maketitle

\tableofcontents
\section{Introduction}
Gravitational-wave~(GW) astronomy paved a new way to look at the Universe, and provides a unique and exquisite probe of black holes (BHs) in the strong-field regime of General Relativity~(GR)~\cite{LIGOScientific:2016lio,Cardoso:2019rvt,LIGOScientific:2021sio,LIGOScientific:2025rid,Berti:2025hly,Cardoso:2025npr}. The coming decades will witness higher sensitivity detectors and hence more powerful tests of our understanding of gravity~\cite{LISA:2017pwj,Branchesi:2023mws}.

Considerable effort has been placed in understanding the two-body problem in GR: the evolution of two gravitationally bound compact objects in vacuum. This problem is assumed to be a good description of the late stages in the inspiral of two BHs or neutron stars. However, the cosmos is filled with matter, and the paradigm that compact objects or even require others to form compact binaries has been strengthening over the years. These would come under the form of a relativistic three-body problem, including so-called binary extreme mass-ratio systems or genuine triple systems of similar-mass components~\cite{Dittmann:2023sha,Yang:2024tje,LIGOScientific:2025brd,Santos:2025ass}. 

The putative presence of companions around evolving compact binaries challenges us to build precise GW waveforms in the presence of matter, a gargantuan effort~\cite{Cardoso:2025npr}, which must start by understanding the basic physics of bodies immersed in a tidal environment. Newtonian results are well known and have indeed been used to constraint the presence of unseen BHs at the center of the Milky Way~\cite{Naoz:2019sjx,Will:2023nlt,GRAVITY:2023met}.
Our purpose here is to study tidal effects in truly strong-field regions, extending already existing results for non-spinning BHs~\cite{Cardoso:2021qqu,Camilloni:2023rra} and for slowly rotating BHs~\cite{Poisson:2014gka,Landry:2015zfa,Pani:2015hfa}. We also aim to take this challenge as an opportunity, by studying the stability of particle motion around tidally deformed spinning BHs, and the impact of companions on the structural evolution of accretion disks and extreme mass-ratio inspirals~(EMRIs).

Throughout this paper, our analysis focuses on timelike bound geodesic motion as an idealized description of particle dynamics around a tidally deformed BH. In realistic EMRIs, the orbital separation evolves due to radiation reaction and therefore the motion departs from a strictly geodesic description~\cite{Mino:1996nk,Quinn:1996am,Poisson:2003nc,Poisson:2011nh}. Likewise, realistic disk dynamics is influenced by additional physical effects, including surrounding magnetic fields, pressure forces, magnetized plasma effects, turbulence, and dissipative processes~\cite{1991ApJ...376..214B,1991ApJ...376..223H,1992ApJ...400..595H,1992ApJ...400..610B,1998RvMP...70....1B,Abramowicz:2011xu}. Our setup is therefore idealized and our results are not directly translated into robust predictions for realistic astrophysical phenomena. Nevertheless, we expect that our geodesics analysis captures important aspects of the underlying phase-space structure and kinematical features of particle motion around a tidally deformed BH. It would be worthwhile to extend our analysis beyond a geodesic description incorporating realistic effects.

This paper is organized as follows. Section~\ref{sec:TidallyDeformedBHs} introduces the metric of a tidally deformed, rotating BH in an adiabatically evolving binary and is devoted to studying the resulting geometry. In Section~\ref{sec:geodesicsinphasespace}, we investigate the phase-space structure of bound timelike geodesics around the deformed BH. Then, in Section~\ref{sec:GeodesisEvolution}, we analyze the structural evolution of initially bound geodesics as the tidal strength is increased. In Section~\ref{sec:AstrophysicalImplications}, we discuss astrophysical implications of our results, particularly focusing on accretion-disk dynamics, EMRIs, and observational impact on GW signals. Section~\ref{sec:Summary} summarizes this work. Appendices~\ref{Sec:tidalperturbation},~\ref{Sec:metricreconstruction}, and~\ref{sec:DeformedSchwarzschild} provide an overview of tidal perturbation theory, metric-reconstruction techniques, and the expressions for the metric of tidally deformed BHs. We employ geometric units, in which Newton's constant and the speed of light $G=c=1$.

\section{Rotating BHs in a binary system}\label{sec:TidallyDeformedBHs}

\subsection{Tidal imprints on BHs}\label{sec:TidalImprintsonBHs}
Stationary, asymptotically flat BHs in vacuum GR must be uniquely described by the Kerr solution~\cite{Robinson:1975bv,Carter:1971zc,Hawking:1971vc,Bekenstein:1996pn,Carter:1997im,Chrusciel:2012jk,Cardoso:2016ryw}. In advanced Kerr coordinates~$(v,r,\theta,\varphi)$, the Kerr metric is given by
\begin{align}
\label{Kerrmetric}
  & g_{\mu\nu}^{(0)}dx^\mu dx^\nu=-\left(1-\frac{2Mr}{\Sigma}\right)dv^2+2dvdr\nonumber\\
  & -\frac{4Mr}{\Sigma} a \sin^2\theta dvd\varphi-2a \sin^2\theta drd\varphi \\
    &+\Sigma d\theta^2+\left(r^2+a^2+\frac{2Mr}{\Sigma}a^2 \sin^2\theta\right)\sin^2\theta d\varphi^2,\nonumber
\end{align}
where $M$ and $a$ represent the Arnowitt–Deser–Misner mass and the spin parameter, respectively; $\Sigma:=r^2+a^2\cos^2\theta$. 
The Kerr BH possesses an event horizon and an inner horizon, located at  $r_\pm:=M\pm\left(M^2-a^2\right)^{1/2}$, respectively. In this work, we assume $M> |a|$.

Now, we place an external, weak gravitational field around the rotating BH. Assuming that the tidal field varies in time slowly compared to the orbital timescale, the tidal interaction with the BH is well described within linear, stationary tidal perturbation theory~\cite{Poisson:2004cw,Taylor:2008xy,Chatziioannou:2012gq}. The tidally deformed metric can be reconstructed from the Weyl scalar that is obtained by solving the Teukolsky equation, following the manner elucidated in Appendices~\ref{Sec:tidalperturbation} and~\ref{Sec:metricreconstruction}. The resulting metric schematically takes the form of   
\begin{align}
    g_{\mu\nu}=g_{\mu\nu}^{(0)}+h_{\mu\nu},\label{eq:metric}
\end{align}
where $h_{\mu\nu}$ is the reconstructed metric. We provide the generator that generates a solution for a given spin parameter online~\cite{Notebook,CoG} and the expression for the components of $h_{\mu\nu}$ in Appendix~\ref{Sec:metricreconstruction}. 

The magnitude of $h_{\mu\nu}$ is characterized by a symmetric and tracefree spatial tensor~${\cal E}_{ij}~(i,j=1,2,3)$, referred to as the quadrupolar gravitoelectric tidal moment, which encapsulates information about the tidal environment. In the present work, we consider a non-precessing, quasi-circular binary at large separation, in which the components of ${\cal E}_{ij}$ with a post-Newtonian expansion satisfy~\cite{Taylor:2008xy}
\begin{align}
    {\cal E}_{11}+{\cal E}_{22}=&-\frac{M_{\rm ext}}{b^3}\left[1+{\cal O}\left(M_{\rm T}/b\right)\right],\label{eq:E11pE22}\\
    {\cal E}_{11}-{\cal E}_{22}=&-\frac{3M_{\rm ext}}{b^3}\left[1+{\cal {O}}\left(M_{\rm T}/b\right)\right]\cos 2\Omega t,\label{eq:E11E22}\\
    {\cal E}_{12}=& -\frac{3M_{\rm ext}}{2b^3}\left[1 +{\cal O}\left(M_{\rm T}/b\right)\right]\sin 2\Omega t,\label{eq:E12}
\end{align}
where $t$ parametrizes the worldline of a tidally deformed body in an ambient weak tidal environment. Here, $M_{\rm ext}$ is mass of the external field; $M_T$ is the total mass of the system, $M_T=M+M_{\rm ext}$; $b~(\gg M_{T})$ is the orbital separation; 
$
    \Omega:=\sqrt{{M_{\rm T}}/{b^3}} \left[1+{\cal O}\left(M_T/b\right)\right]
    $ 
is the orbital angular velocity. Although the tidally deformed metric apparently depends on $t$ through ${\cal E}_{ij}$, the geometry is assumed to be stationary within static tides. 

It is worth emphasizing that the (adiabatic) orbital motion introduces a $t$ - and $\varphi$ - dependence of $h_{\mu\nu}$ only through the phase in the form of $\sqrt{M/b^3}t-\varphi$,~(see, e.g., Eq.~\eqref{eq:TidalStrength}). Here, $t$ denotes a worldline parameter that labels the family of instantaneous geometries in a tidal environment; it should not be interpreted as a physical coordinate time. Henceforth, we set $t=0$ without loss of generality, thereby analyzing an instantaneous snapshot of a deformed geometry in the presence of an external tidal field at $(\theta,\varphi)=(\pi/2,0)$. The subsequent sequence describing the evolving binary at $t=0+\Delta t$ is then generated by the phase shift $\varphi\to \varphi-\sqrt{M/b^3} \Delta t$ from the configuration at $t=0$. The ``time'' dependence is suppressed by $b$, which is much larger than $M$, and hence, the effect of the binary evolution on the subsequent discussion is not significant.

\begin{figure}
\centering
 \includegraphics[scale=0.80]{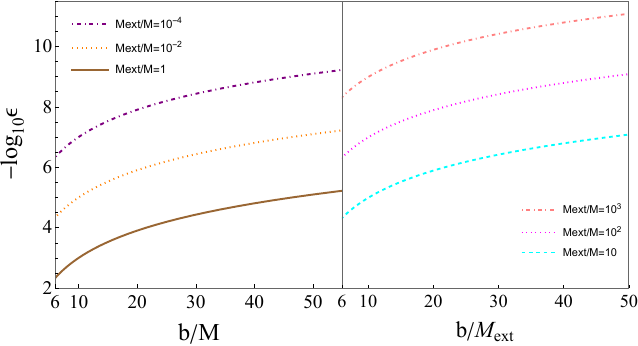}

\caption{Magnitude of $\epsilon$ as a function of the orbital separation $b/M$, $b/M_{\rm ext}$, for various mass ratios.
}
\label{fig:Epsilon}
\end{figure}

We introduce a dimensionless, non-negative parameter,
\be
\epsilon= \frac{M^2 M_{\rm ext}}{b^3}\,,\label{eq:epsilon}
\ee
to count the order of the tidal perturbation. In the following analysis, we perform a perturbative expansion in terms of $\epsilon$ to linear order. The parameter $\epsilon$ is much smaller than unity, as the validity of the post-Newtonian expansion for Eq.~\eqref{eq:E11E22} requires $b\gtrsim \max[6M,6M_{\rm ext}]$, which yields
\begin{align}
    \epsilon \lesssim \frac{1}{216}\times 
    \begin{cases}\dfrac{M_{\rm ext}}{M},\quad\qquad (M\ge M_{\rm ext}),\\
\left(\dfrac{M}{M_{\rm ext}}\right)^2,\quad (M\le M_{\rm ext}).\label{eq:PNrequirement}
    \end{cases}
\end{align}
The typical magnitude of $\epsilon$ is shown in Fig.~\ref{fig:Epsilon} across various mass ratios. 

\subsection{Geometry of deformed BHs}
Stationarity is preserved within static tides. Furthermore, the reflective symmetry under $\theta\to \pi -\theta$ remains within our framework. The deformed geometry, however, is no longer axisymmetric. Moreover, the spacetime no longer belongs to Petrov~type D spacetime~\cite{Kinnersley:1968zz}, losing hidden symmetry of the unperturbed Kerr metric. The components of the resulting metric increase with $r$, indicating that the deformed geometry is not asymptotically flat. It is worth noting that the validity of the deformed metric extends only up to $r\ll (b^3/M_{\rm ext})^{1/2}$, or equivalently to $r/r_+\ll 1/\epsilon^{1/2}$~\cite{Poisson:2009qj}.

\begin{figure}
\centering
   \includegraphics[scale=0.45]{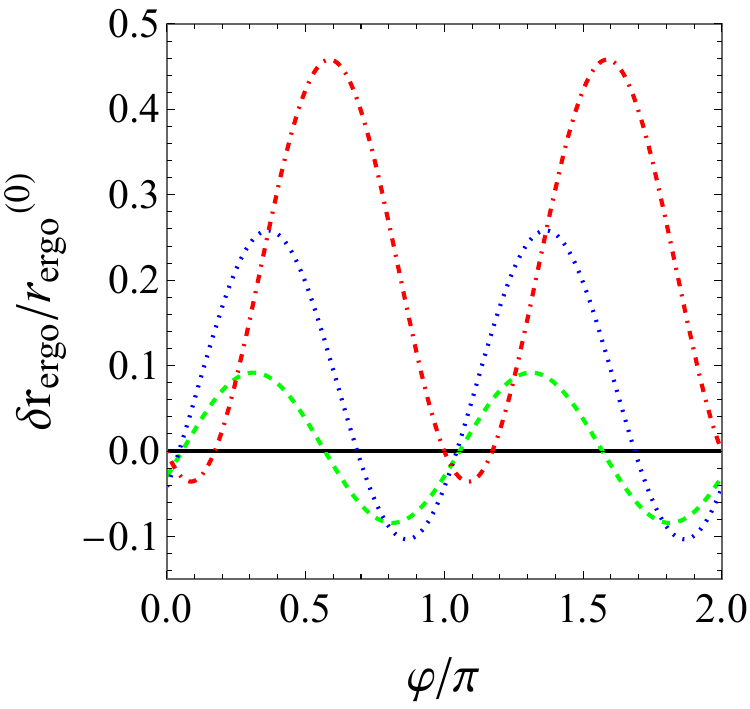}
\caption{$\epsilon$-independent deviation of the radius of the static-limit surface from $r_{\rm ergo}^{(0)}:=M+\sqrt{M^2-a^2\cos^2\theta}$ for the BH with $a/M= 4/5$, cf. Eq.~\eqref{eq:rergo_shift}. Colors denote $\theta=0,\pi$~(black solid), $\theta=0.1\pi,0.9\pi$~(green dashed),~$\theta=0.25\pi,0.75\pi$~(blue dotted), and $\theta=0.5\pi$~(red dot-dashed), which reflects the symmetry with respect to the equatorial plane.}
\label{fig:Geometry}
\end{figure}

In the current coordinate system, we find a null hypersurface at $r=r_+^{\cal E}$, where $r_+^{\cal E}$ satisfies $g^{rr}|_{r=r_+^{\cal E}}={\cal O}(\epsilon^2)$, given by 
\begin{align}
    r_+^{\cal E}= r_++{\cal O}\left(\epsilon^2\right).
\end{align}
This expression implies that $r_+^{\cal E}$ coincides with the radius of the event horizon of the unperturbed BH within the linear perturbation. However, the notion of an ``event horizon'' in the deformed geometry is not accurate, as the perturbed metric describes only the local spacetime.

The local geometry of a given surface~$(v,r)=({\rm const.},{\rm const}.)$ is characterized by null congruences across the surface. The expansion scalars,~$\Theta$ and $\Theta_{(-)}$, of future-directed, outgoing and ingoing null congruences with tangent vector fields,~${\bf K}$ and ${\bf N}$, such that $g_{\mu\nu}K^\mu N^\nu =-1+{\cal O}(\epsilon^2)$, are, respectively, determined by
$
    \Theta:= \gamma^{\mu\nu}\nabla_\mu K_\nu
$
and  
$
\Theta_{(-)}:= \gamma^{\mu\nu}\nabla_\mu N_\nu,
$
where $\gamma^{\mu\nu}=K^\mu N^\nu+N^\mu K^\nu + g^{\mu\nu}$ with $\gamma_{\mu\nu}K^\nu=0=\gamma_{\mu\nu}N^\nu$. The induced metric,~$\gamma_{\mu\nu}$, depends on $\varphi$ in addition to $\theta$, implying deformations of the shape of two-dimensional surface at $(v,r)=({\rm const},{\rm const}.)$ due to the presence of tidal fields. We then find
\begin{align}
&\Theta \big|_{r < r_+^{\cal E}} < 0, \quad \Theta \big|_{r=r_+^{\cal E}}={\cal O}(\epsilon^2),\quad \Theta \big|_{r_+^{\cal E}<r\ll r_+/\epsilon^{1/2}}>0,\\
&\Theta_{(-)}|_{r<r_+^{\cal E}}<0, \quad \Theta_{(-)}|_{r=r_+^{\cal E}}<0, \quad \Theta_{(-)}|_{r_+^{\cal E}<r\ll r_+/\epsilon^{1/2}}<0.\nonumber
\end{align}
It follows that the surface~$r=r_+^{\cal E}$ is foliated by apparent horizons -- the outermost marginally outer trapped surface on a given time slice -- within the current approximation. These results extend previous analysis of the non-rotating case in Ref.~\cite{Poisson:2009qj} to the rotating case. Although the intrinsic geometry of the two-dimensional surface at $r=r_+^{\epsilon}$ for a given~$v$ becomes angle-dependent, the coordinate location of the apparent horizon remains unchanged. Another intriguing observation is that both $\Theta$ and $\Theta_{(-)}$ can change sign at large distances around $r\sim r_+/\epsilon^{1/2}$, depending on $\theta$ and $\varphi$. Although the present perturbative description is no longer accurate in this regime, it would be worth exploring this asymptotic behavior in a follow-up analysis.

Static observers with four-velocity parallel to the time-translation Killing vector field cannot exist in the ergoregion, enclosed by the static-limit surface at $r=r_{\rm ergo}$ where $g_{vv}|_{r=r_{\rm ergo}}={\cal O}(\epsilon^2)$. This surface of the deformed geometry deviates from that of the unperturbed Kerr BH:
\begin{align}
    r_{\rm ergo}=M+\sqrt{M^2-a^2\cos^2\theta}+\epsilon \delta r_{\rm ergo}\left(\theta,\varphi\right)+{\cal O}\left(\epsilon^2\right).\label{eq:rergo_shift}
\end{align}
The relative error in the radii of the ergoregion of the unperturbed spacetime and of the perturbed spacetime is of order of, at most, ${\cal O}(10)\times \epsilon~\%$, as displayed in Fig.~\ref{fig:Geometry}. This feature is insensitive to the BH spin.

It is worth emphasizing that the geometric deformations observed above are indeed of physical significance. BHs in vacuum GR are believed to exhibit vanishing tidal Love numbers~(TLNs), implying that tidal fields induce no higher multipole moments of BHs~\cite{Binnington:2009bb,Chia:2020yla,Charalambous:2021mea,Poisson:2021yau}. The values of relativistic TLNs depend on the chosen definition of tidal and induced multipole moments, i.e., a subjective decomposition of a perturbed metric, without a robust matching scheme with GW signals, and are therefore ambiguous~(see Refs.~\cite{Gralla:2017djj,Poisson:2020vap,Katagiri:2024wbg,Katagiri:2024fpn}). Crucially, a tidally deformed metric does deviate from the unperturbed one and is uniquely determined once an appropriate boundary condition is imposed~\cite{Katagiri:2024wbg,Katagiri:2024fpn}. This ensures that the geometric properties, particle motion, and GW propagation in tidally deformed spacetimes are entirely free of ambiguities.

\section{Bound geodesics in phase space}\label{sec:geodesicsinphasespace}

\subsection{System and setup}
We analyze motion of timelike bound geodesics whose tangent vector~${\bf u}$ satisfies
\begin{align}
     u^\mu \nabla_\mu u^\nu=0.\label{eq:GeodesicEquation}
\end{align}
The vector field,~${\bf u}$, is normalized by $g_{\mu\nu}u^\mu u^\nu=-1$  within linear order of $\epsilon$ and the components are expressed as $u^\mu=(dv/d\lambda,dr/d\lambda,d\theta/d\lambda,d\varphi/d\lambda)$ with the affine parameter~$\lambda$. In the unperturbed Kerr spacetime, Eq.~\eqref{eq:GeodesicEquation} is integrable by virtue of three conserved quantities associated with the Killing symmetries  -- energy~$\hat{E}:=-g_{\mu\nu}(\partial/\partial v)^\mu u^\nu$, angular momentum~$\hat{L}:=g_{\mu\nu} (\partial/\partial\varphi)^\mu u^\nu$, and the Carter constant -- in addition to the conservation of rest mass~\cite{Bardeen:1972fi}. This property ensures separability of variables of Eq.~\eqref{eq:GeodesicEquation} and regular motion. By contrast, in the tidally deformed Kerr spacetime, constants of motion are insufficient, and Eq.~\eqref{eq:GeodesicEquation} no longer decouples, which generically gives rise to chaotic motion~\cite{2002ocda.book.....C,Gair:2007kr,Lukes-Gerakopoulos:2012qpc,Destounis:2025tjn}. 

We numerically solve Eq.~\eqref{eq:GeodesicEquation} with the initial condition,~
\be
(r,dr/d\lambda,\theta,\varphi)=(r_0,0,\pi/2,0)\,.\label{eq:InitialData}
\ee
The initial data at $\lambda=0$,  together with $\hat{E}$, $\hat{L}|_{\lambda=0}$, and the rest mass conservation, determine the other initial parameters,~$(dv/d\lambda,d\theta/d\lambda,d\varphi/d\lambda)|_{\lambda=0}$.\footnote{It is not necessary to specify $v|_{\lambda=0}$ within our current analysis, as the Christoffel symbols and the geodesic equation in a stationary spacetime are explicitly independent of the time coordinate.}The specific energy, ${\hat{E}}$, is used to monitor numerical accuracy, while the evolution of $\hat{L}$ indicates the degree to which axisymmetry is broken. 

We set the upper limit of the integration of Eq.~\eqref{eq:GeodesicEquation} so as to satisfactorily capture the behavior of bound orbits in the presence of a fixed tidal field. To this end, we compare two relevant timescales: the orbital timescale of the bound geodesics and the timescale associated with the variation of the tidal field. First, assuming that bound orbits approximately follow Kepler's law and their typical radius is of order of $M$, the timescale for $n$ cycles is estimated as
\begin{align}
    T_{\rm geod}\sim 2\pi n M.
\end{align}
Next, Eqs.~\eqref{eq:E11pE22}--\eqref{eq:E12} imply that the characteristic timescale of the variation of the tidal field is
\begin{align}
    T_{\rm tides}\sim 2\pi \sqrt{\frac{M_{\rm ext}}{M+M_{\rm ext}}}\frac{M}{\sqrt{\epsilon}}.
\end{align}
We now require that the configuration of the tidal field remains approximately unchanged during $n$ cycles, i.e., $T_{\rm geod}\ll T_{\rm tides}$, yielding 
\begin{align}
n\ll \sqrt{\frac{M_{\rm ext}}{M+M_{\rm ext}}}\frac{1}{\sqrt{\epsilon}}.
\end{align}
Based on the above analysis, we integrate Eq.~\eqref{eq:GeodesicEquation} up to $\lambda=3\times 10^5r_+$, which is sufficient to obtain orbits with $680-1000$ cycles, for various initial radii~$r_0$.

\subsection{Non-integrability, resonance, and chaos }
\begin{figure*}
\centering
 \includegraphics[scale=0.232]{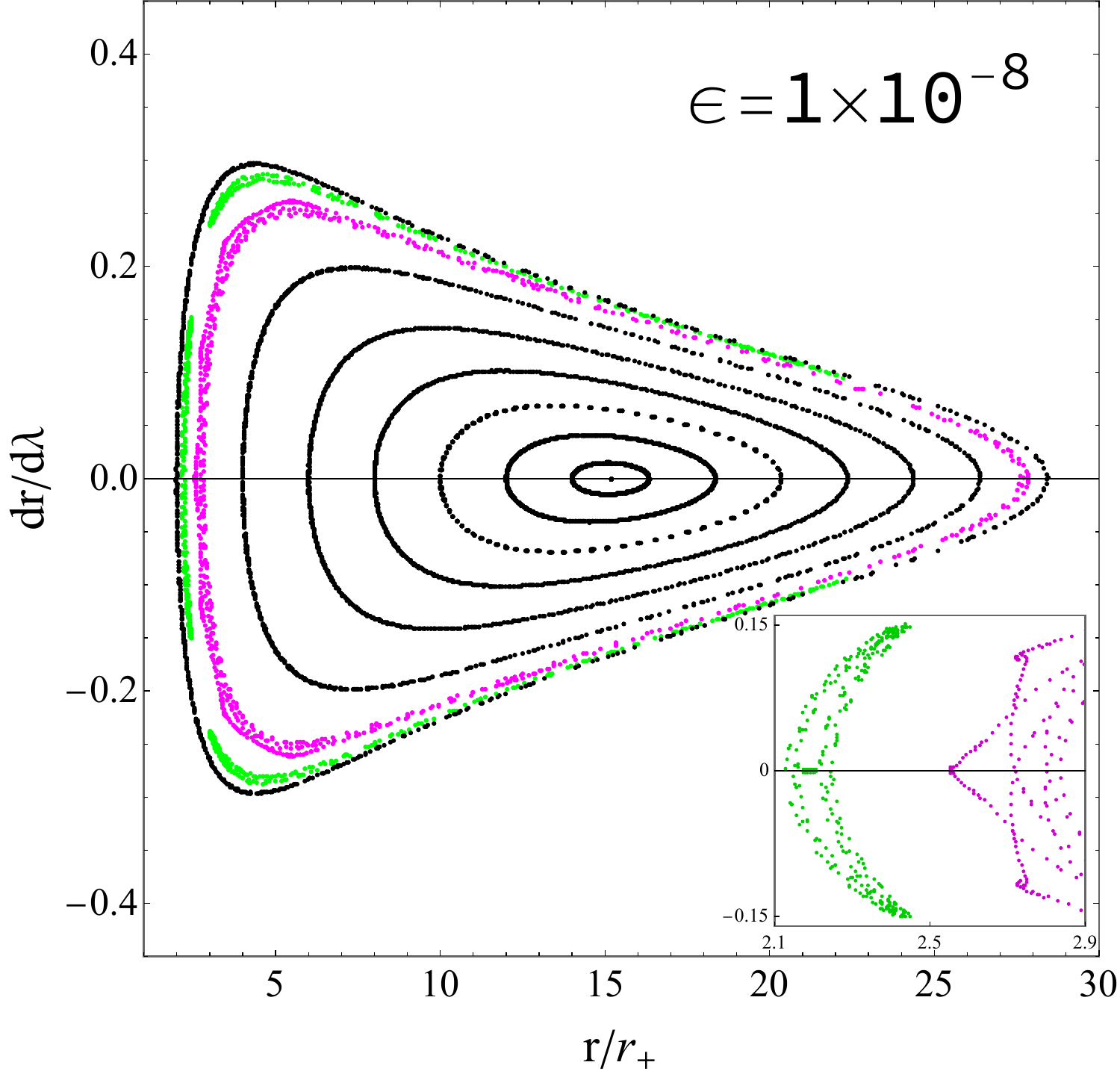}
  \includegraphics[scale=0.232]{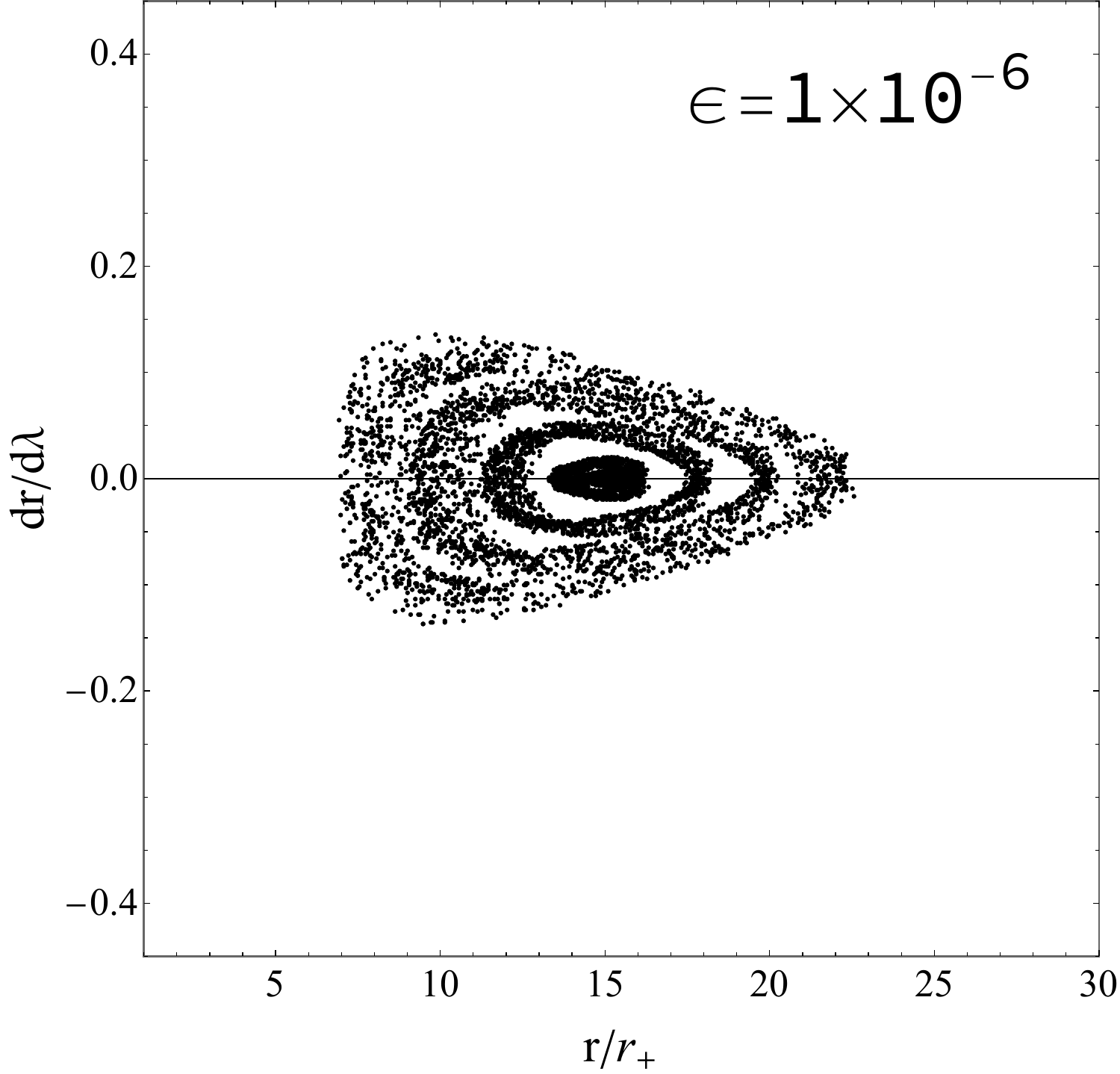}
  \includegraphics[scale=0.232]{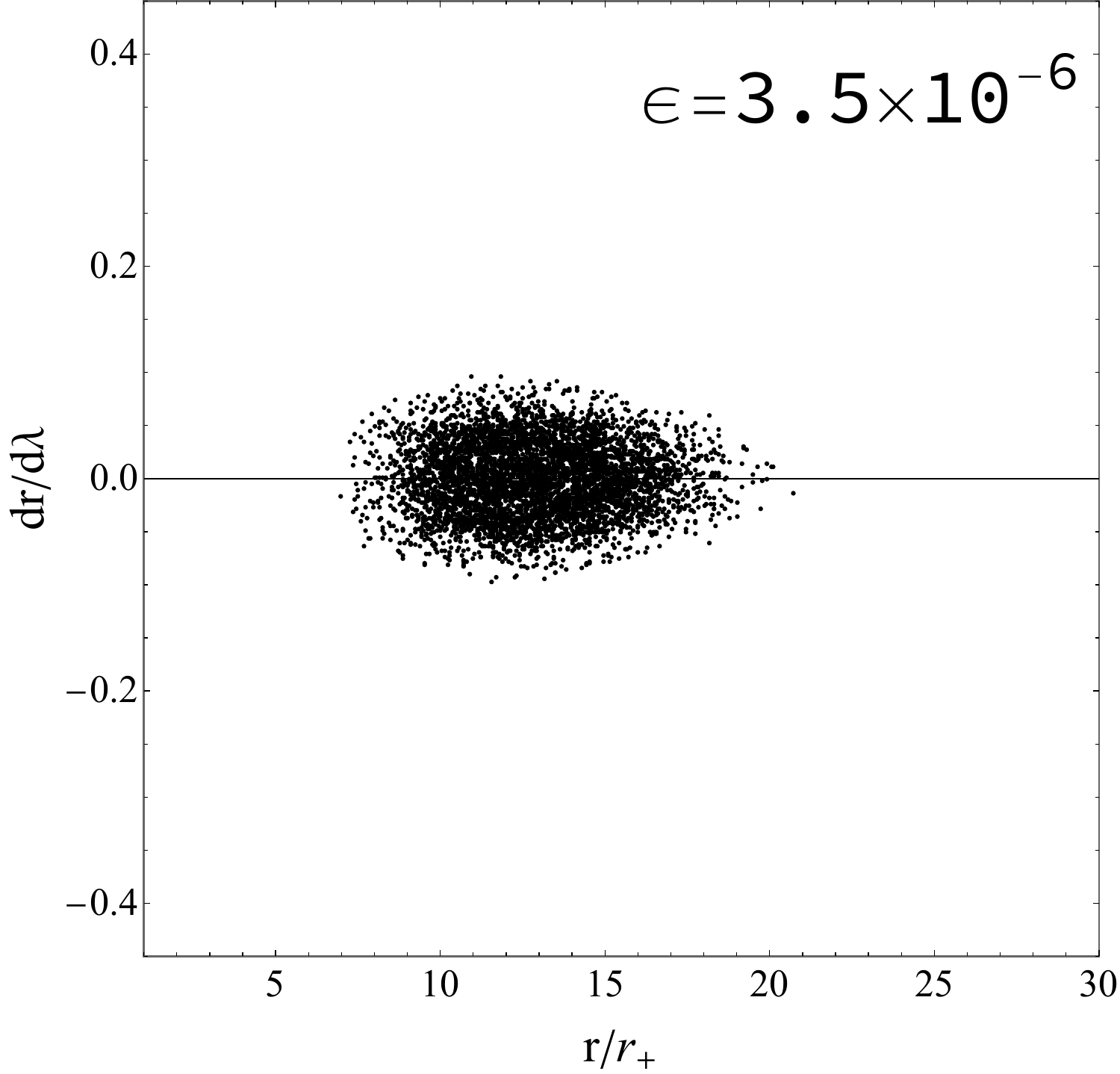}
\caption{Prograde bound orbits with initial parameters $(\hat{E}, \hat{L}|_{\lambda=0})=(0.98,1.8r_+)$ which starts its motion at various discrete values of $r_0$, around a tidally deformed Kerr BH with $a/M=4/5$. In the left panel, the sequences in green and magenta correspond to representative examples of resonant orbits with frequency ratios~$2/3$ and $5/7$, respectively. Their initial radii are chosen near those of the corresponding resonant orbits in the unperturbed Kerr case. The inset shows an enlarged figure of the detailed resonant structure. The dot at $(d\lambda/dr,r/r_+)\simeq (0,15.176)$ in the left panel corresponds to an originally spherical orbit at $r=r_{\rm sph}$ in Eq.~\eqref{eq:rsph}.} 
\label{fig:PoincareMap}
\end{figure*}
Bound geodesic motion around a Kerr BH is effectively equivalent to geodesic flow on a two-dimensional invariant torus wrapped by an orbit with $2$ degrees of freedom in four-dimensional phase space arising from the reduction of the Hamiltonian using the constants of motion~(see Fig.~1 of Ref.~\cite{Brink:2015roa}). Each orbit on an invariant torus is characterized by two fundamental frequencies, $\omega_r$ and $\omega_\theta$, associated with radial and polar motion, respectively~\cite{Schmidt:2002qk,Brink:2008xy}. The azimuthal fundamental frequency,~$\omega_\varphi$, then characterizes the azimuthal drift of the orbit along the torus. The combination,~${\bm k}\cdot{\bm \omega}:=p\omega_r-q\omega_\theta-m \omega_\varphi$ with relatively prime integers~$p$ and $q~(<p)$ and the azimuthal number~$m$, classifies orbits as non-resonant or resonant. Specifically, when ${\bm k}\cdot{\bm \omega}\neq0$~(non-resonant), the corresponding orbit is quasi-periodic, covering the torus densely. By contrast, when ${\bm k}\cdot{\bm \omega}=0$~(resonant), the frequencies are commensurate and the motion no longer densely fills the torus.

A system that deviates from Kerr bound geodesics perturbatively, in general, loses its integrability. According to the Kolmogorov-Arnold-Moser~(KAM) theorem~\cite{Arnold:1989who}, invariant tori associated with ``sufficiently irrational'' orbits remain stable, albeit slightly deformed, under sufficiently small perturbations. In other words, modest perturbations to an integrable system do not immediately destroy most of the invariant tori~\cite{1993CeMDA..56..191L}. The condition for a ``sufficiently irrational'' frequency ratio is given by~\cite{2002ocda.book.....C}
\begin{align}
    \left|{\bm k}\cdot{\bm \omega}\right|> \frac{{\cal K}}{\left(p+q+|m|\right)^3},\label{eq:DiophantosCondition}
\end{align}
where ${\cal K}$ is a small numerical factor that depends on the nature of perturbations. Larger values of $p+q+|m|$ and/or a smaller value of ${\cal K}$ tend to satisfy Eq.~\eqref{eq:DiophantosCondition}; otherwise, the stability of the corresponding tori is no longer guaranteed, indicating that these tori are more easily destroyed.

Moreover, the Poincar{\'e}-Birkhoff theorem~\cite{Poincar1912SurUT,Birkoff1913} states that resonant tori are generically destroyed by perturbations, typically giving rise to a set of $p$~stable periodic orbits and $p$~unstable periodic orbits for the tori with $\omega_r/\omega_\theta=q/p$, in the region where the tori initially occupied~\cite{Lukes-Gerakopoulos:2010ipp,Lukes-Gerakopoulos:2012qpc}. The stable orbits generate ``islands'' of regular motion -- often referred to as resonant islands -- in their vicinity, where quasi-periodic orbits survive with a frequency ratio close to $q/p$ over time evolution. By contrast, the unstable orbits generate a pair of stable and unstable asymptotic manifolds, i.e., families of orbits with different initial conditions that, respectively, approach or repel the unstable orbit. Orbits that wander outside the resonant islands and intersect these manifolds become highly sensitive to initial conditions and exhibit complicated patterns of motion, thereby giving rise to chaotic layers surrounding the islands. The destruction of resonant tori is a clear manifestation of the non-integrability of perturbed or deformed Kerr systems~\cite{2002ocda.book.....C,Gair:2007kr,Apostolatos:2009vu,Lukes-Gerakopoulos:2010ipp,Lukes-Gerakopoulos:2012qpc,Destounis:2025tjn}. 

As the amplitude of the perturbation increases, the deformed non-resonant tori begin to break down. At the same time, the chaotic layers generated by different resonances expand and may eventually overlap~\cite{2002ocda.book.....C}. Chaotic signatures arising from a single resonance are no longer separated each other but become global. The dynamics is then dominated by chaos, which will be demonstrated in the next section.

It is worth noting that there is a hierarchy among resonances. Low-order resonances with small values of~$q+p$, such as $(q,p)=(1,2)$, $(1,3)$, or $(2,3)$ typically emerge close to the innermost stable circular orbit~(ISCO) for unperturbed BHs in a manner insensitive to the details of non-integrable perturbations, whereas higher-order resonances tend to lie in the weak-field regime~\cite{Brink:2013nna,Brink:2015roa}. The $(2,3)$ resonance is of particular astrophysical interest because quadrupolar perturbations preferentially excite resonances with $q=2$, and hence, the resonance with $p=3$ is expected to be prominent~\cite{Apostolatos:2009vu,Lukes-Gerakopoulos:2010ipp,Brink:2013nna,Brink:2015roa}. 

\subsection{Poincar{\'e} surface of section}

A section constructed by recording the intersections of phase orbits with a chosen plane -- known as a Poincar{\'e} surface of section~\cite{Levin:2008mq,Apostolatos:2009vu,Lukes-Gerakopoulos:2010ipp} -- reveals the phase-space structure of bound geodesics in the $(r,dr/d\lambda)$ plane. A quasi-periodic orbit corresponds to a closed curve on the surface, while a periodic orbit appears as a finite set of discrete points with the number of points determined by the winding number per cycle. Non-regular orbits emerge as a finite set of randomly scattered points.

In what follows, we analyze the Poincar{\'e} surface of section chosen on the equatorial plane~($\theta=\pi/2$) with representative values of $\epsilon$. We restrict ourselves to $a/M=4/5$ but the results presented in the following are qualitatively insensitive to $a$. 

In the unperturbed Kerr spacetime, non-resonant orbits that start with different values of $r_0$ appear as a set of regular closed curves around a spherical orbit, while resonant orbits emerge as a finite set of discrete points around it~\cite{Gueron:2001ex,Gair:2007kr,Levin:2008mq,Apostolatos:2009vu,Lukes-Gerakopoulos:2010ipp}. The radius of spherical prograde orbits,~$r_{\rm sph}$, around the Kerr BH is determined by~\cite{Teo:2020sey}
\begin{equation}
\begin{split}
        \hat{E}=&\frac{r_{\rm sph}^3\left(r_{\rm sph}-2M\right)-a\left(a\hat{Q}-\sqrt{\Upsilon}\right)}{r_{\rm sph}^2\sqrt{r_{\rm sph}^3\left(r_{\rm sph}-3M\right)-2a\left(a\hat{Q}-\sqrt{\Upsilon}\right)}},\label{eq:rsph}\\
    \hat{L}=&-\frac{2M a r_{\rm sph}^3+\left(r_{\rm sph}^2+a^2\right)\left(a\hat{Q}-\sqrt{\Upsilon}\right)}{r_{\rm sph}^2\sqrt{r_{\rm sph}^3\left(r_{\rm sph}-3M\right)-2a\left(a\hat{Q}-\sqrt{\Upsilon}\right)}},
\end{split}
\end{equation}
where $\Upsilon:=M r_{\rm sph}^5-\hat{Q}(r_{\rm sph}-3M)r_{\rm sph}^3+a^2 \hat{Q}^2$ with the Carter constant,~$\hat{Q}=\Sigma^2(u^\theta)^2+\cos^2\theta [a^2(1-\hat{E}^2)+\hat{L}^2/\sin^2\theta]$~\cite{Carter:1968rr}. The radius, $r_{\rm sph}$, increases with increasing $a/M$ and/or $\hat{E}$. For instance, $r_{\rm sph}/r_+\simeq 15.176$ when $(\hat{E},\hat{L})=(0.98,1.8r_+)$ with $a/M=4/5$ and $\hat{Q}\simeq 7.34r_+^2$.

\begin{figure}
\centering
 \includegraphics[scale=0.33]{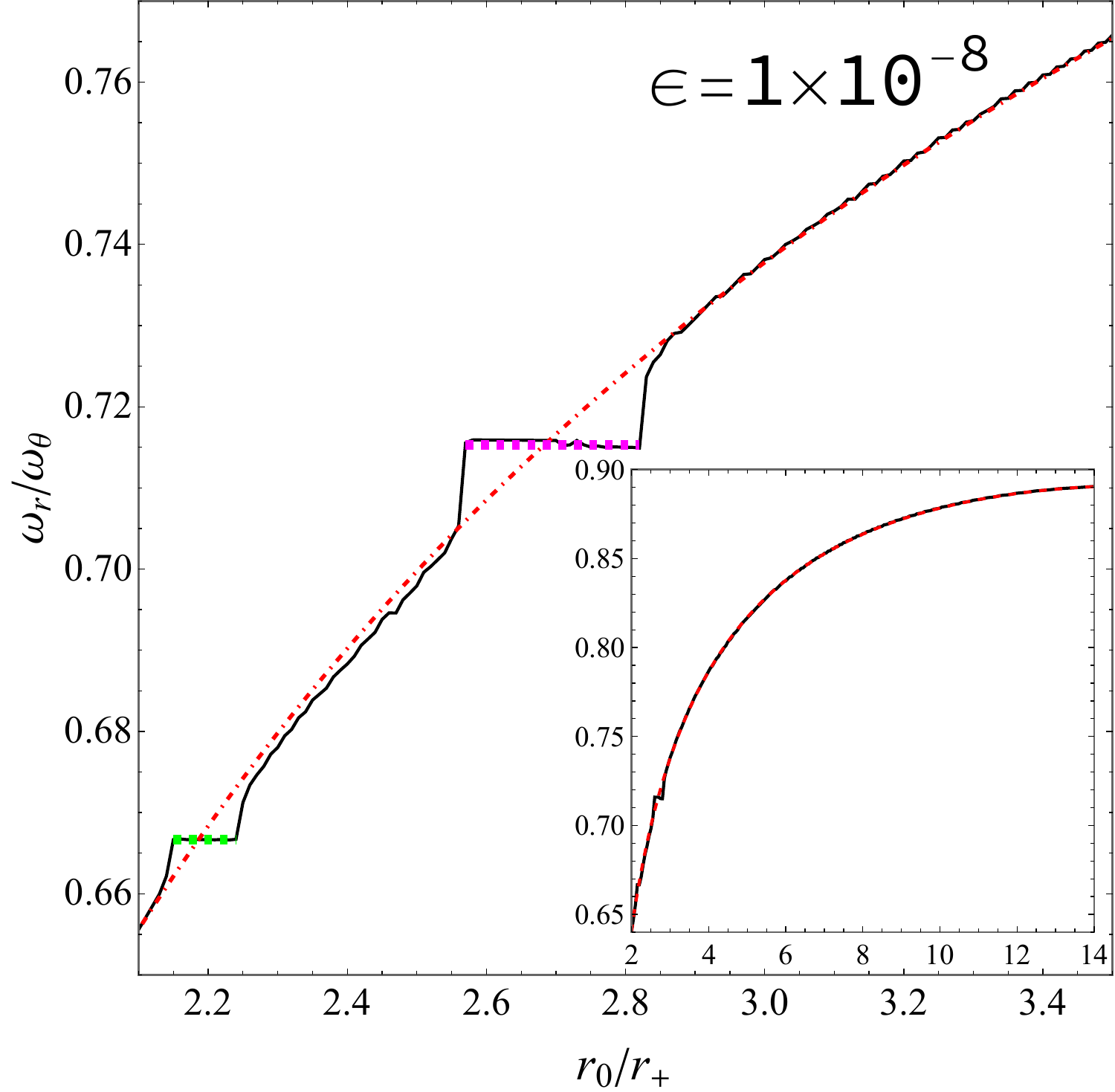}
\caption{Frequency ratio, ~$\omega_r/\omega_\theta$, associated with the $(q,p)=(2,3)$~(green) and $(5,7)$~(magenta) resonant orbits, highlighted in the corresponding color in the left panel of Fig.~\ref{fig:PoincareMap}. The red dot-dashed smooth curve denotes the unperturbed Kerr case computed semi-analytically following the procedure of Refs.~\cite{Schmidt:2002qk,Fujita:2009bp,Brink:2013nna,Brink:2015roa}. The inset presents the overview figure, showing that no clear plateaus are visible outside the enlarged region.}
\label{fig:FreqRatio}
\end{figure}

The left panel of Fig.~\ref{fig:PoincareMap} shows that, in the tidally deformed Kerr spacetime with $\epsilon=10^{-8}$, most of the orbits still form closed curves around $(r,dr/d\lambda)=(r_{\rm sph},0)$ on the equatorial Poincar{\'e} surface of section up to $680$ orbital cycles, being consistent with the KAM theorem~\cite{Arnold:1989who}. 
It is worth emphasizing that the resonant orbits
with frequency ratios~$2/3$ and $5/7$ are highlighted in green and magenta, respectively. Their initial radii are chosen near those of the corresponding resonant orbits in the unperturbed Kerr case. In the inset, the green structure is more suggestive for a stable (periodic) resonant orbit in its core, whereas the magenta structure appears as a thinner and more obscured layer, potentially due to accumulation of numerical errors. Although the latter does not resolve into a clear island structure, its localization and correspondence with the $5/7$ frequency ratio indicate that it originates from resonance.  Indeed, we find that orbits in the vicinity of these representative orbits share nearly the same rational frequency ratio, as shown in Fig.~\ref{fig:FreqRatio}. Here, we estimate $\omega_r/\omega_\theta$ from the average intervals in $\lambda$ between successive points at which $dr/d\lambda$ and $\theta-\pi/2$ change sign from negative to positive.\footnote{Our approach differs from methods based on the rotation angle, as employed in the literature, e.g., Refs.~\cite{Apostolatos:2009vu,Lukes-Gerakopoulos:2012qpc,Destounis:2023gpw}. Nevertheless, we obtain consistent results also with the rotation-angle approach. Moreover, in the unperturbed Kerr case, we have confirmed that the frequency ratio extracted from the fully numerical long-term geodesics integrations agree well with the semi-analytic prediction obtained following the procedure of Refs.~\cite{Schmidt:2002qk,Fujita:2009bp,Brink:2013nna,Brink:2015roa}.} We also find that $\omega_r/\omega_\varphi$ and $\omega_\theta/\omega_\varphi$ exhibit plateaus at nearly rational numbers within the same range of $r_0$ as in Fig.~\ref{fig:FreqRatio}; the ratios imply $(\omega_r/\omega_\theta,\omega_r/\omega_\varphi,\omega_\theta/\omega_\varphi)=(2/3,3/5,9/10), (5/7,2/3,14/15)$, indicating resonances with $(q,p,m)=(2,3,0),(5,7,0)\cdots$.\footnote{Within our present setup -- a non-precessing, quasi-circular binary system -- the tidal perturbation  contains no $m=\pm 1$ contributions.} It is worth emphasizing that the width of the plateaus depends on the magnitude of $\epsilon$ but does not directly measure the resonance strength; instead, it indicates the parameter interval over which the orbit stays near resonances.

As $\epsilon$ increases, the outer curves on the equatorial Poincar{\'e} surface of section begin to collapse, higher-resonant tori are destroyed, and thus chaotic regions increasingly expand, as illustrated in the middle panel of Fig.~\ref{fig:PoincareMap}. Bound orbits are found in the range of $6.9\lesssim r_0/r_+\lesssim 23.8$ up to $680$ orbital cycles, which is smaller than the range of the unperturbed case, $1.7\lesssim r_0/r_+ \lesssim 28.7$. The originally spherical orbit persists, albeit with slight deformation. We find that the originally bound orbits that start outside this range plunge into the BH, as demonstrated in Section~\ref{sec:GeodesisEvolution}. Chaos becomes the dominant feature of the dynamics. No clear plateaus are found in $\omega_r/\omega_\theta$ over different values of $r_0$; instead, irregular, jagged variations are observed.

Further increasing $\epsilon$ enhances the destruction of non-resonant tori and leads to overlapping of the chaotic layers arising from different resonances, thereby expanding chaotic regions. The right panel of Fig.~\ref{fig:PoincareMap} demonstrates that, in the deformed Kerr spacetime with $\epsilon=3.5\times 10^{-6}$, bound orbits in $11.4 \lesssim r_0/r_+ \lesssim 17.8$ spread throughout the phase space. In other words, the tidal perturbation destroys all the originally (quasi-)periodic orbits including the spherical orbit. Clear plateaus with a rational number are not found in $\omega_r/\omega_\theta$ in the corresponding range of $r_0$. We find unbound orbits, as well as plunging and bound orbits, appearing in an apparently random fashion, as demonstrated in Section~\ref{sec:GeodesisEvolution}.

\section{Geodesics in evolving binary}\label{sec:GeodesisEvolution}

\subsection{Orbits in configuration space}\label{sec:GeodesisinConfigurationSpace}
%
\begin{figure*}
\centering
  \includegraphics[scale=0.32]{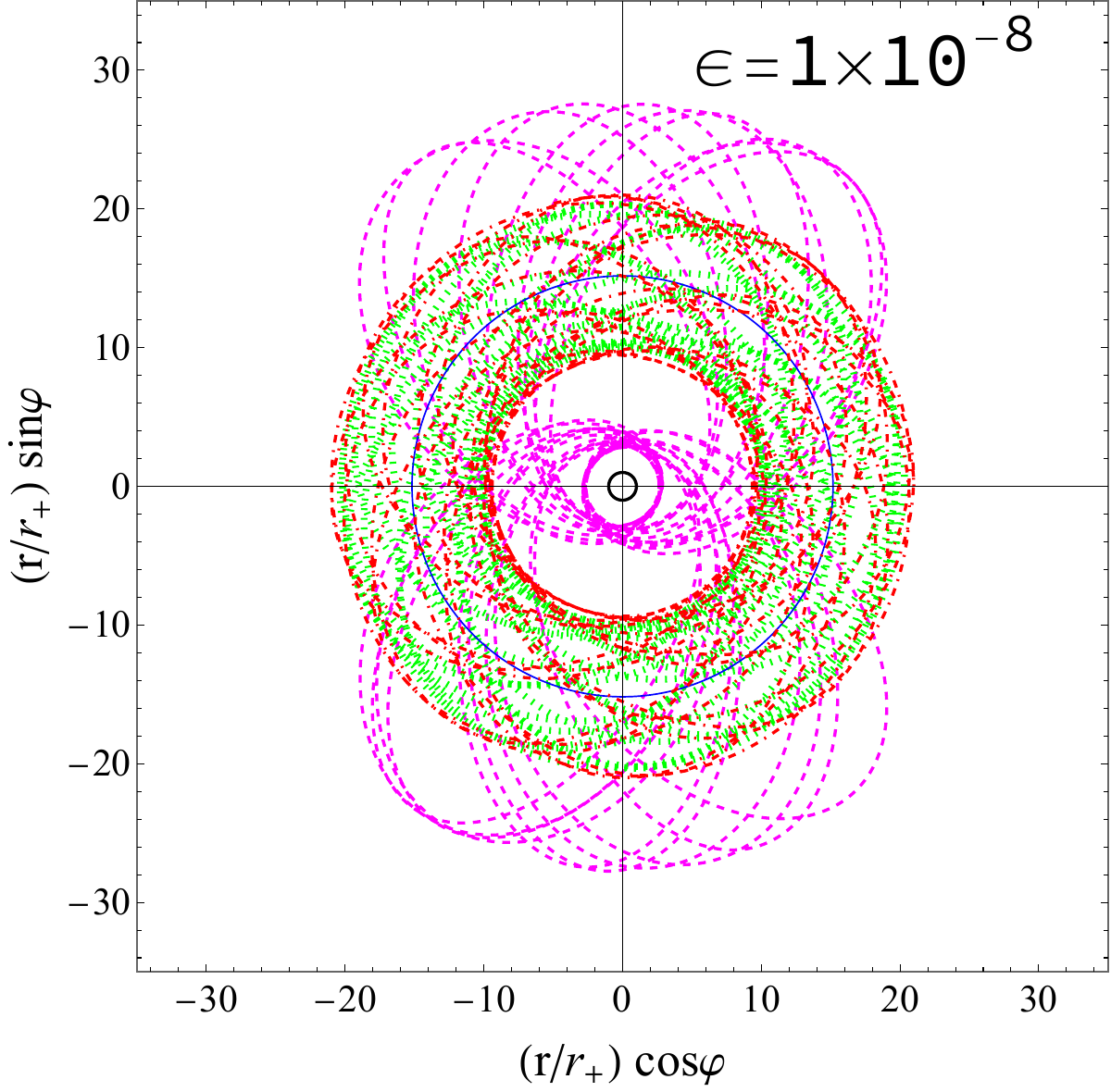}
   \includegraphics[scale=0.32]{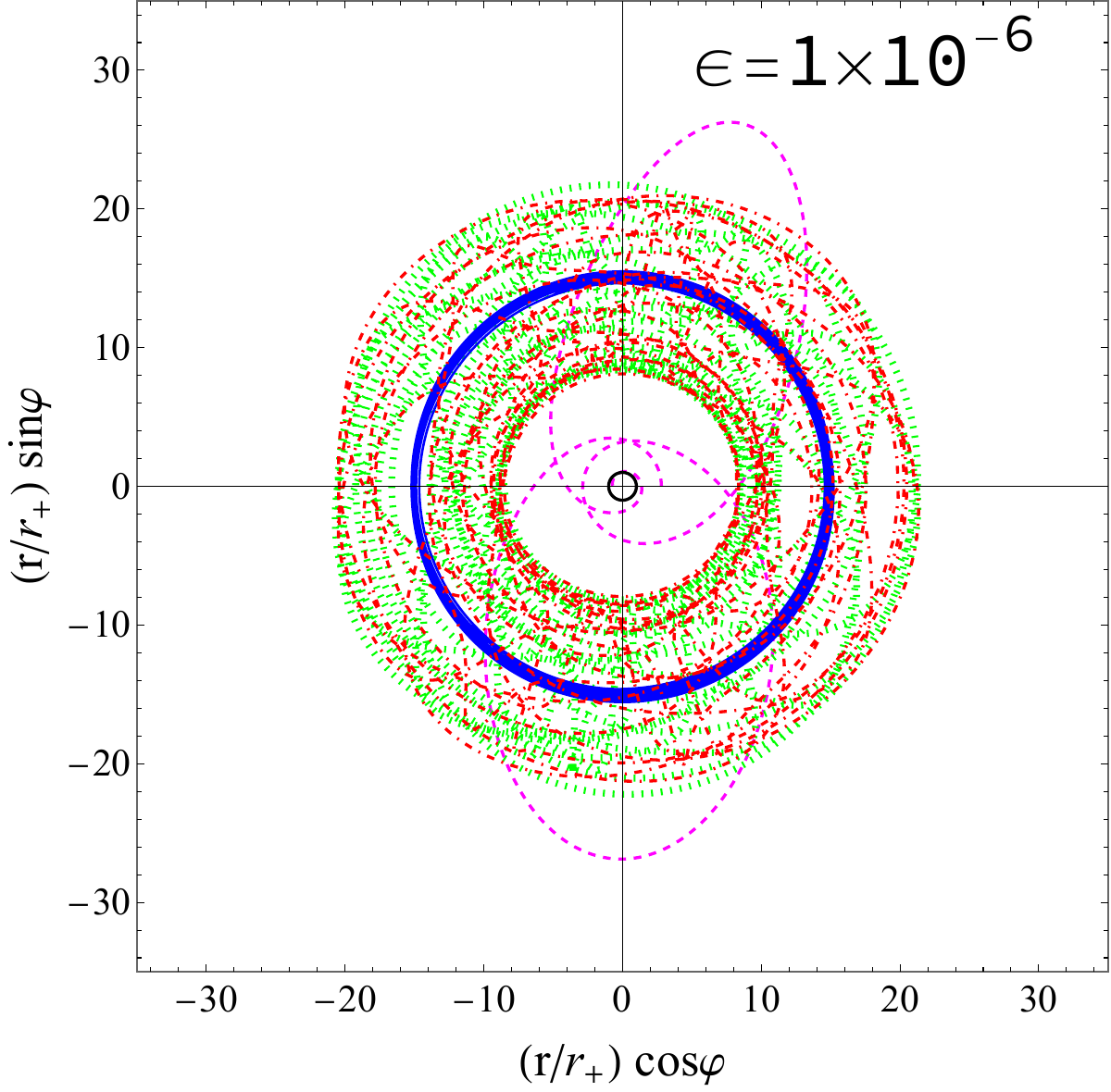}
  \includegraphics[scale=0.32]{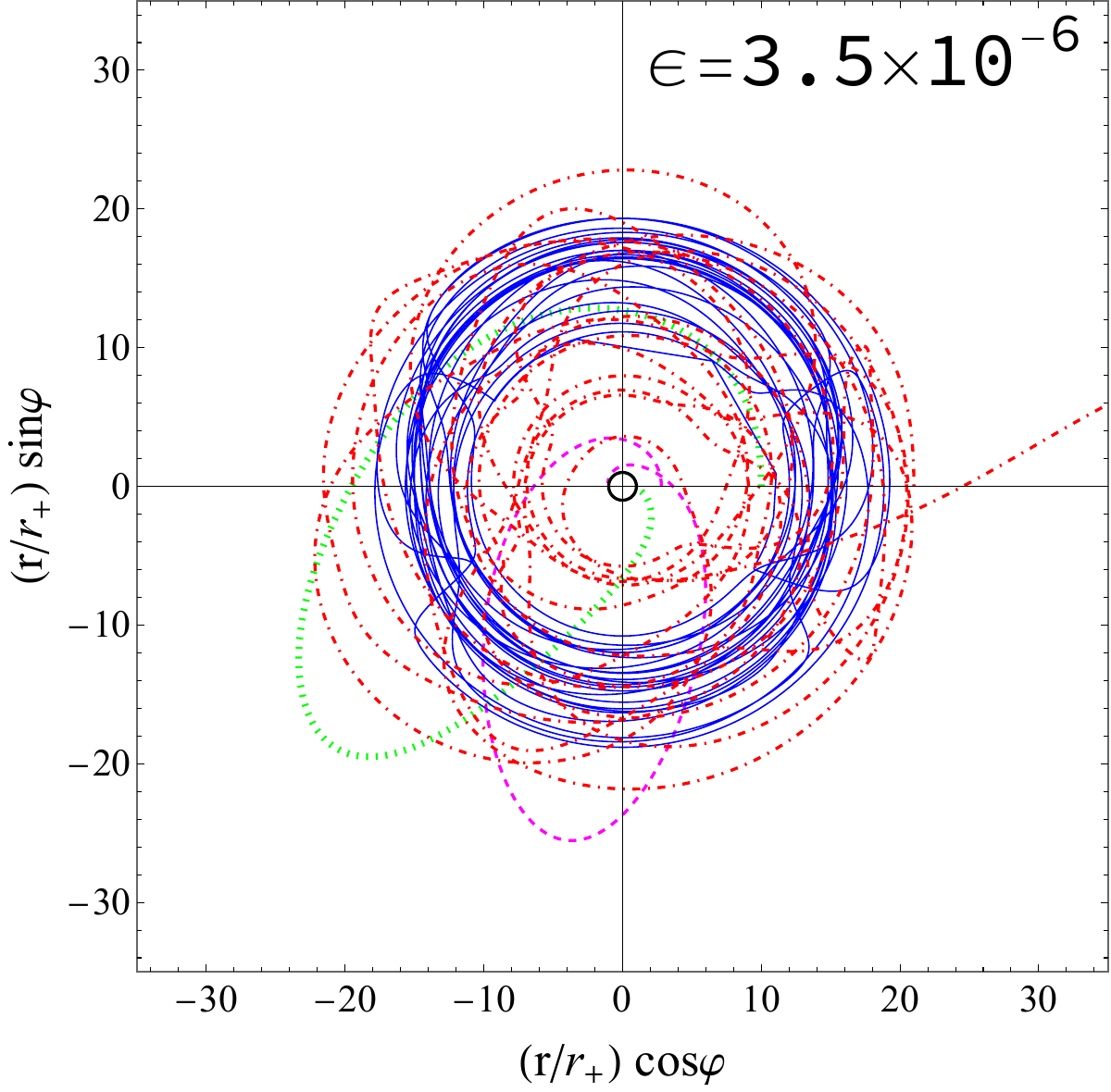}
  \includegraphics[scale=0.32]{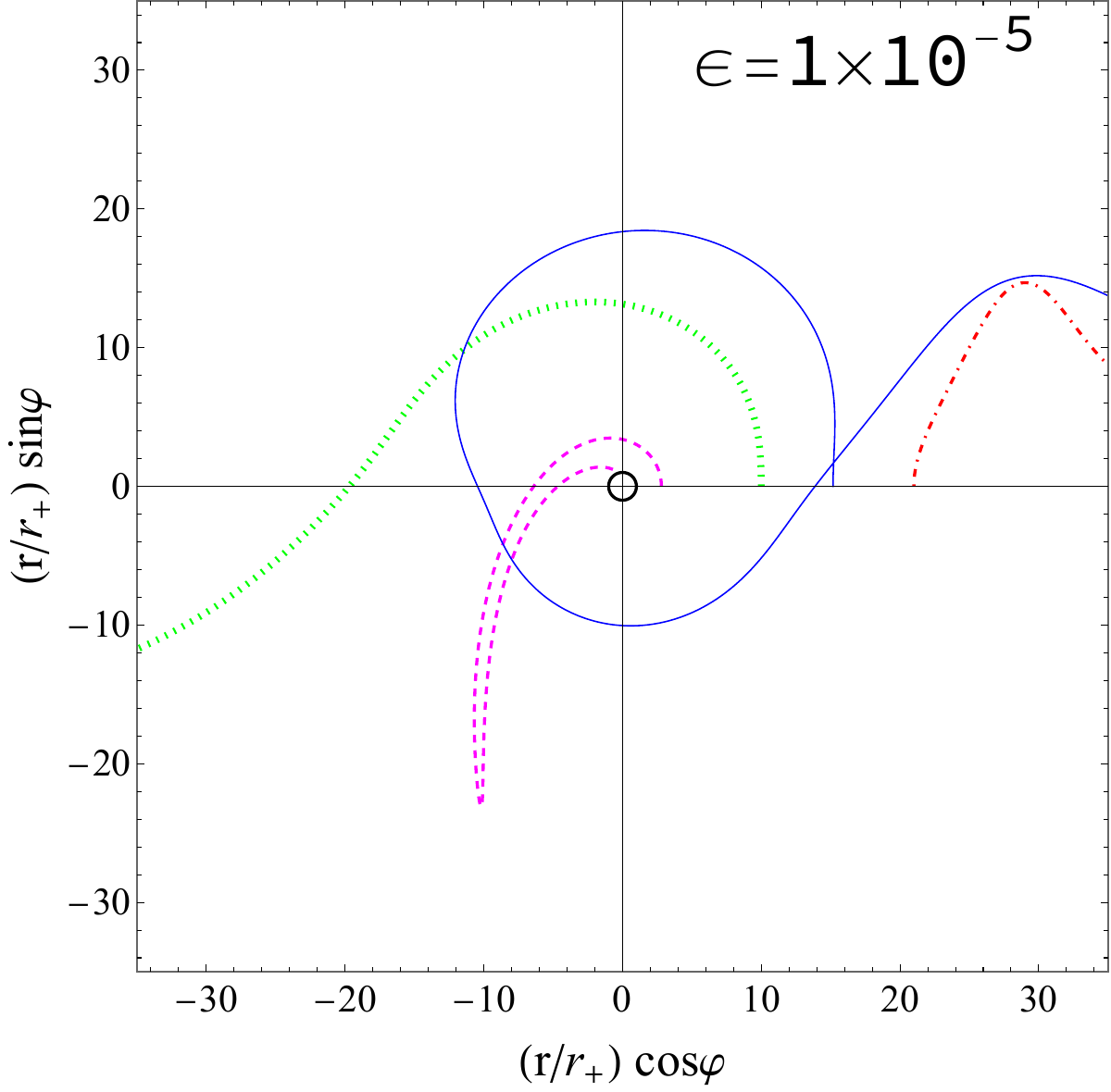}
\caption{Evolution of originally bound orbits with initial parameters $(\hat{E},
\hat{L}|_{\lambda=0})=(0.98,1.8r_+)$, evolved from $r_0/r_+=2.8$~(magenta dashed), $r_0/r_+=10$~(green dotted), $r_0/r_+=15.176$~(blue solid), $r_0/r_+=21$~(red dot-dashed), in the tidally deformed Kerr spacetime with $a/M=4/5$ for various values of $\epsilon$. Note that the blue solid trajectories correspond to an originally spherical orbit. Top-left, top-right, bottom-left, and bottom-right panels represent bound phase, chaotic phase, collapsing phase, and depleted phase, defined in the main text, respectively.}
\label{fig:EvolutionOfConfig}
\end{figure*}
%
%
We discuss the evolution of bound geodesics following Eq.~\eqref{eq:GeodesicEquation} with various values of $\epsilon$ in configuration space. As representatively illustrated in Fig.~\ref{fig:EvolutionOfConfig}, we find typical four distinct phases separated by three critical values of $\epsilon$, i.e., $\epsilon_c^-$, $\epsilon_c^+$, and $\epsilon_c^0$, in the evolution at fixed ($\hat{E},\hat{L}|_{\lambda=0}$) as follows:
\begin{itemize}
\item {\bf Bound phase}~($0\le \epsilon <\epsilon_c^-$): All originally bound orbits that start with a different value of $r_0$ remain bound around the deformed BH. The geodesic structure largely retains its regular character.

\item {\bf Chaotic phase}~($\epsilon_c^-\le \epsilon <\epsilon_c^+$): Originally bound orbits begin to collapse to plunging orbits into the BH but most of the bound orbits persist with chaos. The chaotic feature tends to be more pronounced, as $\epsilon$ increases. 

\item {\bf Collapsing phase}~($\epsilon_c^+\le \epsilon < \epsilon_c^0$): Unbound orbits escaping in the directions of $\varphi=0,\pi$ appear, in addition to bound and plunging orbits.

\item {\bf Depleted phase}~($\epsilon\ge \epsilon_c^0$):  Bound orbits eventually disappear entirely, leaving plunging and unbound orbits.
\end{itemize}
For smaller values of $\hat{E}$, all bound orbits tend to collapse to plunging orbits without unbound orbits emerging, corresponding to a direct transition from chaotic phase to depleted phase. As will be discussed in the next section, the emergence of plunging orbits can be thought of as a result of loss of $\hat{L}$ due to torques induced by an external tidal field, while unbound orbits arise from quadrupolar deformation stretching
along the direction towards the tidal field with a decreasing effective potential.

It is worth noting that the critical values of $\epsilon$ depend on $\hat{E}$ and $\hat{L}$, as well as $a/M$. For our example of Fig.~\ref{fig:EvolutionOfConfig}, $(\epsilon_c^-,\epsilon_c^+,\epsilon_c^0)\simeq (2\times 10^{-8},2\times 10^{-6},5\times10^{-6})$. It appears that these critical amplitudes tend to increase slightly by less than a factor of $10$ as $a/M$ increases, suggesting that spin effects are subdominant. In other words, for a fixed-mass ratio, the modest change in $\epsilon$ across $a/M$ translates into, at most, an ${\cal O}(10^{1/3})$ variation in the binary separation, which does not affect qualitative conclusions.

\subsection{Critical tidal amplitude}\label{sec:CriticalTidalAmplitude}
We discuss the underlying mechanism of each transition and estimate the typical values of $\epsilon_c^-$, $\epsilon_c^+$, and $\epsilon_c^0$. To this end, we consider an equatorial geodesic motion around a tidally deformed Schwarzschild BH, which is sufficient to provide the estimates for the rotating case with generic orbital inclination.

\subsubsection{Effective potential}
The metric of a tidally deformed Schwarzschild BH is provided in Appendix~\ref{sec:DeformedSchwarzschild}. Following the manner of Ref.~\cite{Cardoso:2021qqu}, we obtain the equation of motion for $u^r$:
\begin{align}
  -  \left(u^r\right)^2+\hat{E}^2-V=0,
\end{align}
where
\begin{align}
    &V=\left(1-\frac{r_+}{r}\right)\left(1+\frac{\hat{L}^2}{r^2}\right)\nonumber\\
    &-\epsilon \left(1+3\cos2\varphi\right)\left(1-\frac{r_+}{r}\right)\label{eq:Potential}\\
    &\times\left[2r\left(r-r_+\right)+\frac{\hat{L}^2}{r^2}\left(4r^2-2r_+r-r_+^2\right)\right]. \nonumber
\end{align}
This recovers Eq.~(23) of Cardoso and Foschi~\cite{Cardoso:2021qqu} by setting $\delta=-1$ and by redefining the bookkeeping parameter therein,~$\epsilon_{\rm CF}$, to $\epsilon_{\rm CF} =-(1/2)(1+3\cos 2\varphi)\epsilon$. 

It follows from Eq.~\eqref{eq:Potential} that $V$ explicitly depends on $\varphi$. The potential, $V$, asymptotically tends
\begin{align}
    &V|_{r\gg r_+}= 1-\frac{r_+}{r}+{\cal O}\left(r_+^2/r^2\right)\nonumber\\
    &-2\left(1+3\cos2 \varphi\right)\epsilon \left[\left(\frac{r}{r_+}\right)^2+{\cal O}\left(r/r_+\right)\right].\label{eq:Potentialatlarger}
\end{align}
As noted before, the tidally deformed metric is valid only up to $r/r_+\ll 1/\epsilon^{1/2}$~\cite{Poisson:2009qj}, which suppresses the contribution at ${\cal O}(\epsilon)$ smaller than unity. It is worth noting that, when $1+3\cos2\varphi>0~(<0)$, $V$ decreases~(increases) with $r$ at large distances. In particular, this trend is more pronounced as $\epsilon$ increases and is most enhanced at $\varphi=0,\pi$~($\varphi=\pi/2,3\pi/2$).

\subsubsection{$\epsilon_c^-$: from bound phase to chaotic phase}

\begin{figure*}
\centering
   \includegraphics[scale=0.285]{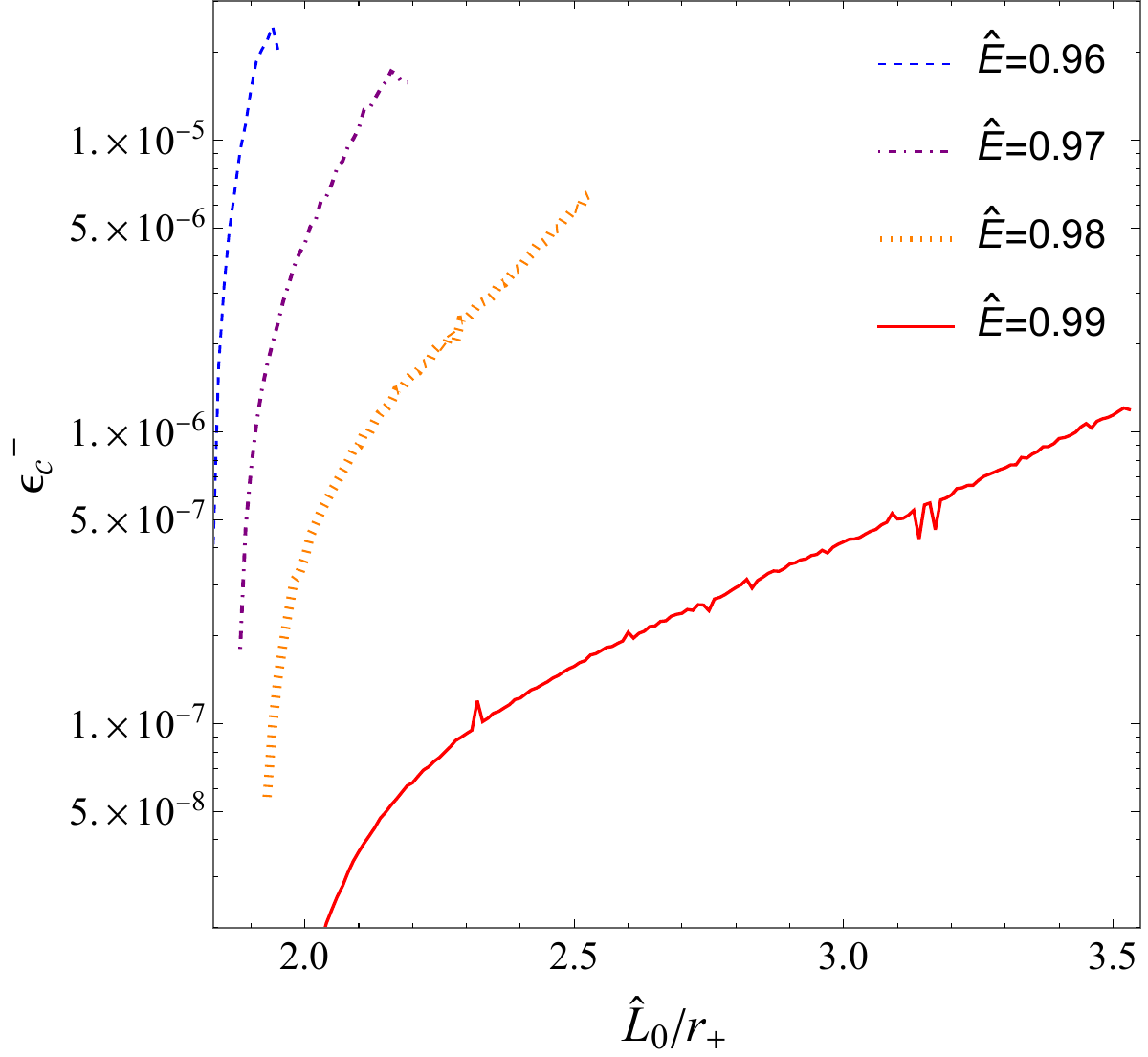}
    \includegraphics[scale=0.285]{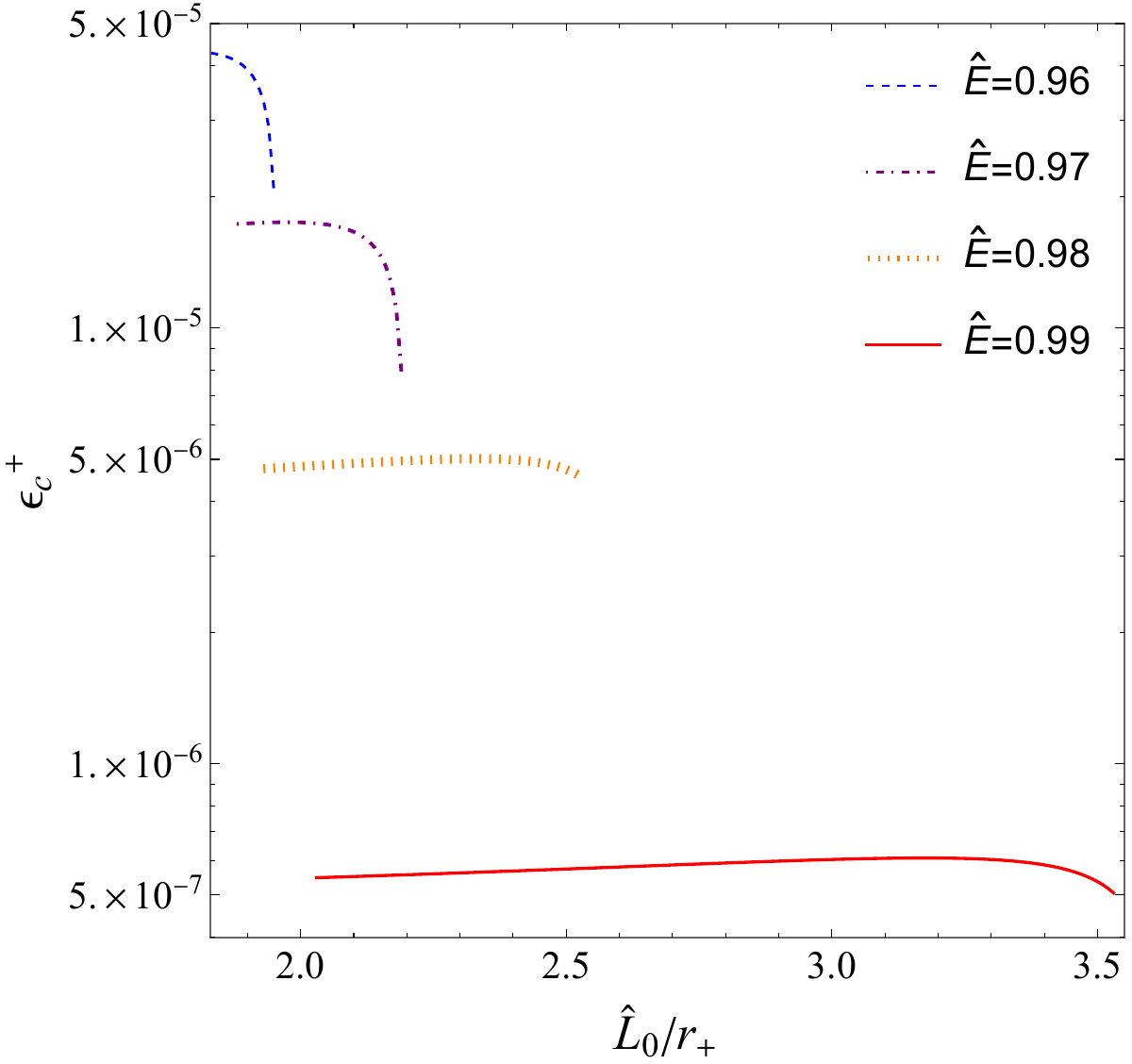}
   \includegraphics[scale=0.285]{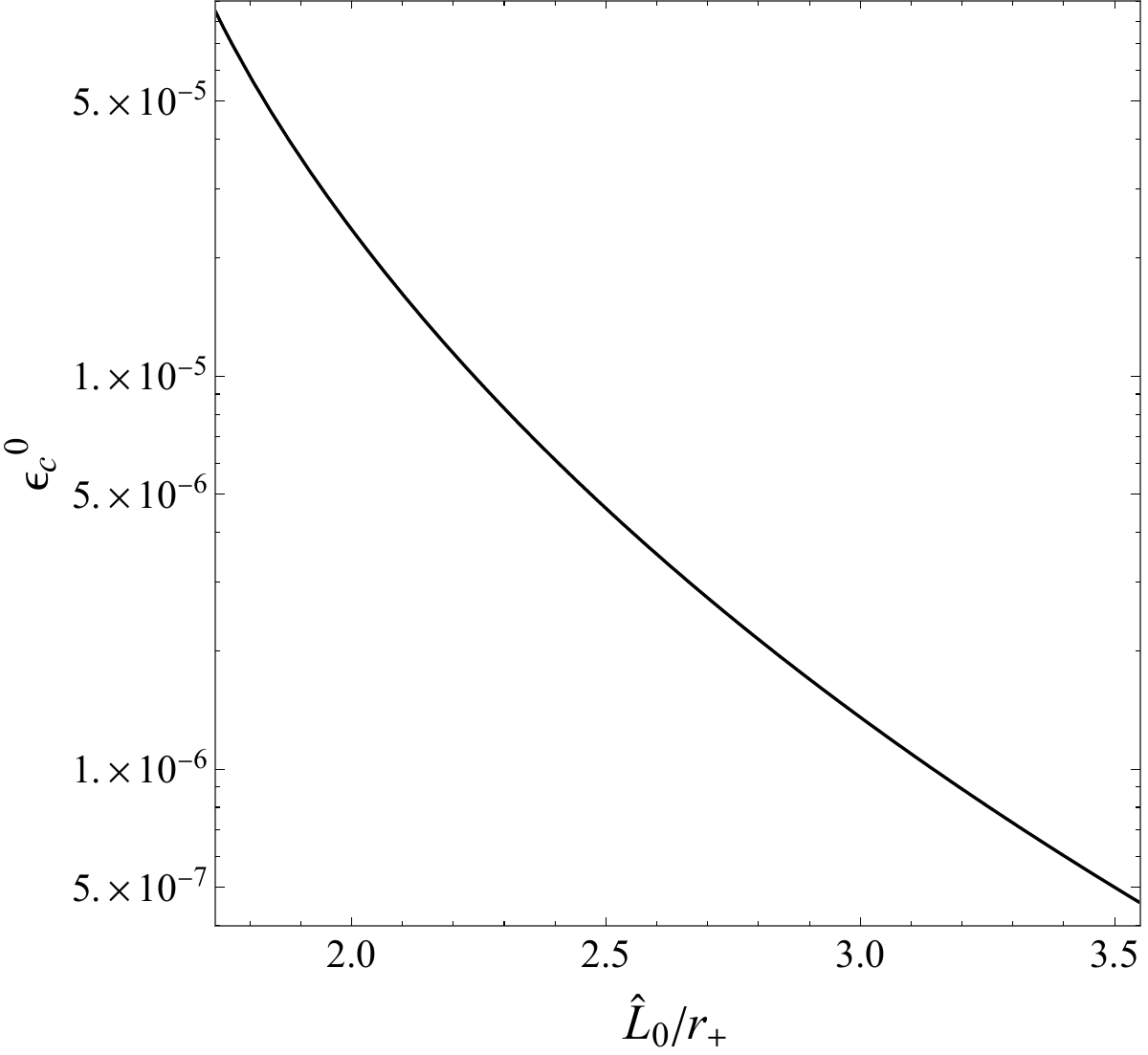}
\caption{{\bf Left:} $\epsilon_c^-$ given in Eq.~\eqref{eq:epsiloncminus} for various sets of $(\hat{E}, \hat{L}_0)$. The curves correspond to $\hat{E}=0.96$~(blue dashed), $\hat{E}=0.97$~(purple dot-dashed), $\hat{E}=0.98$~(orange dotted),~$\hat{E}=0.99$~(red solid). {\bf Middle:} $\epsilon_c^+$ given in Eq.~\eqref{eq:Epsiloncplus} for the same set of $(\hat{E},\hat{L}_0)$ as in the left panel. {\bf Right:} $\epsilon_c^0$ given in Eq.~\eqref{eq:Epsilonc0}. Note that, given $\hat{L}_0$, $\hat{E}$ is determined through Eq.~\eqref{eq:rsph} with $\hat{Q}=0$. }
\label{fig:EpsilonandL0}
\end{figure*}

The plunging behavior primarily arises from the decrease in $\hat{L}$ during the last orbit prior to the plunge, which lowers the potential barrier in the vicinity of the BH horizon. To estimate $\epsilon_c^-$, we analyze the evolution of $\hat{L}$ and $V$. The Lagrangian for a geodesic, ${\cal L}=(1/2)g_{\mu\nu}u^\mu u^\nu$, leads to the Euler-Lagrange equation, 
$d\hat{L}/d\lambda=\partial_\varphi {\cal L}$, thereby obtaining
\begin{align}
   \frac{d\hat{L}}{d\lambda}=6\epsilon \sin2\varphi\left(\frac{r}{r_+}\right)^2\left[1-\frac{r_+}{r}+\frac{2\hat{L}^2}{r^2}\left(1-\frac{r_+}{2r}-\frac{r_+^2}{4r^2}\right)\right].\label{eq:EvolutionOfLhat}
\end{align}
This result implies that the tidal field induces torques, arising from the relative motion between the geodesic and the tidal field at $\varphi=0$, thereby reducing $\hat{L}$ when $\pi/2<\varphi<\pi$. Within the perturbative framework (small $\epsilon$), the variables on the right-hand side in Eq.~\eqref{eq:EvolutionOfLhat} can be replaced with those in the unperturbed Schwarzschild spacetime. Therefore, $\hat{L}$ reads 
\begin{align}
     \hat{L}=\hat{L}_0+\epsilon \delta \hat{L},
\end{align} 
with
\begin{align}
&\hat{L}_0:=\hat{L}_{\rm Sch}\left(=\hat{L}\big|_{\lambda=0}\right),\\
&\delta \hat{L}= 6 \int_0^{\lambda_{\rm pl}} d\lambda~ \sin2\varphi_{\rm Sch}\left(\frac{r_{\rm Sch}}{r_+}\right)^2\label{eq:deltaLinlambda}\\
    &\qquad\times\left[1-\frac{r_+}{r_{\rm Sch}}+\frac{2\hat{L}_0^2}{r_{\rm Sch}^2}\left(1-\frac{r_+}{2r_{\rm Sch}}-\frac{r_+^2}{4r_{\rm Sch}^2}\right)\right]\nonumber,
\end{align}
where $\lambda_{\rm pl}$ is the value of the affine parameter at the onset of plunge and the subscript,~``${\rm Sch}$'', indicates the unperturbed quantities. Henceforth, we omit the subscript for the variables.

With $d\lambda/dr=(\hat{E}^2-V)^{-1/2}|_{\epsilon=0}$, Eq.~\eqref{eq:deltaLinlambda} is cast into
\begin{align}
&\delta \hat{L}= 12 \int_{r_{\rm in}}^{r_{\rm out}}  \frac{dr}{\sqrt{\hat{E}^2-V_{\rm Sch}}} ~ \sin2\varphi \left(\frac{r}{r_+}\right)^2\label{eq:deltaLinr}\\
    &\qquad\times\left[1-\frac{r_+}{r}+\frac{2\hat{L}_0^2}{r^2}\left(1-\frac{r_+}{2r}-\frac{r_+^2}{4r^2}\right)\right]\nonumber,\\
   & V_{\rm Sch}:=V\big|_{\epsilon = 0},
\end{align}
where $r_{\rm in/out}$ is the radii of the inner and outer turning points of bound orbits, i.e., pericenter/apocenter, in the unperturbed Schwarzschild spacetime, determined by $\hat{E}^2=V_{\rm Sch}|_{r=r_{\rm in/out}}$, respectively. The angle,~$\varphi$, is expressed as
\begin{align}
    \varphi\left(r\right)=2\int_{r_{\rm in}}^{r} \frac{dr}{\sqrt{\hat{E}^2-V_{\rm Sch}}} \frac{\hat{L}_0}{r^2}+2n \int_{r_{\rm in}}^{r_{\rm out}}  \frac{dr}{\sqrt{\hat{E}^2-V_{\rm Sch}}} \frac{\hat{L}_0}{r^2},
\end{align}
where the second term on the right-hand side accounts for the periastron shift during $n$ cycles.

Equation~\eqref{eq:deltaLinr} allows us to predict $\delta\hat{L}$ once $n$ is specified. To estimate the typical magnitude of $\delta \hat{L}$ that maximizes the reduction of the potential barrier, we search for the most negative value of $\delta \hat{L}$ in Eq.~\eqref{eq:deltaLinr} by varying $n$ from $0$ to $100$. The resulting value of $\delta \hat{L}$ oscillates quasi-periodically with respect to $n$. Its minimum takes a negative value of order of $-\delta \hat{L}/r_+= {\cal O}(10^3)- {\cal O}(10^5)$, depending on $\hat{E}$ and $\hat{L}_0$.

As $\epsilon$ increases, $\hat{L}$ tends to decrease further during evolution while $\hat{E}$ remains constant, leading to plunging orbits for which $\hat{E}^2>V$ holds at the peak location~$r=r_{\rm peak}$. For the unperturbed case, $r_{\rm peak}$ is expressed as
\begin{align}
r_{\rm peak}=\frac{\hat{L}_0^2}{r_+}\left(1-\frac{r_+}{\hat{L}_0}\sqrt{\left(\frac{\hat{L}_0}{r_+}\right)^2-3}\right).\label{eq:rpeak}
\end{align}
Note that $\hat{L}_0>\sqrt{3}r_+$ for originally bound orbits. Equation~\eqref{eq:rpeak} provides a good approximation for the peak location of $V$. 

The condition, $\hat{E}^2>V_{\rm Sch}|_{r=r_{\rm peak}}$ with $\hat{L}=\hat{L}_0+\epsilon \delta\hat{L}$, then sets a lower bound on $\epsilon$, which approximately corresponds to $\epsilon_c^-$,
    \begin{align}
    \epsilon_c^-=-\frac{\hat{L}_0}{2\delta \hat{L}}\left[1+\left(\frac{r_{\rm peak}}{\hat{L}_0}\right)^2\frac{\left(1-\hat{E}^2\right)r_{\rm peak}-r_+}{r_{\rm peak}-r_+}\right].\label{eq:epsiloncminus}
\end{align}
The typical value of $\epsilon_c^-$ is presented in the left panel of Fig.~\ref{fig:EpsilonandL0}. We have verified agreement of the above estimate to a number of full numerical results obtained from Eq.~\eqref{eq:GeodesicEquation} within the correct order of magnitude. 

Finally, we argue that an initially ISCO transitions to plunging orbits in the presence of tidal fields. To see this, we substitute $\hat{L}=\sqrt{3} r_+$ -- the specific angular momentum at the ISCO for the Schwarzschild BH -- into $dV/dr$, yielding
\begin{align}
\label{eq:dVatISCO}
&\frac{dV}{dr}=\frac{\left(r-3r_+\right)^2r_+}{r^4}\\
&-\epsilon\left(1+3\cos2\varphi\right)\frac{4r^5-4r^4r_++18 r^2r_+^3-6rr_+^4-9r_+^5}{r^4r_+^2}.\nonumber
\end{align}
This shows that $dV/dr$ is positive throughout $r>r_+$ in particular directions such that $1+3\cos2\varphi<0$, implying that the orbit is no longer stable or cannot climb the potential well to escape.

\subsubsection{$\epsilon_c^+$: from chaotic phase to collapsing phase}
As Eq.~\eqref{eq:Potentialatlarger} implied before, $V$ decreases at large distances when $1+3\cos 2\varphi>0$, leading to unbound orbits in the particular directions. The transition to collapsing phase can be understood in terms of the properties of $V$ around an outer turning point of bound orbits. In the following analysis, we neglect the change of $\hat{L}$.

We first estimate the radius of an outer turning point,~$r_{\rm out}$, by exploiting $r_{\rm out}$ in the unperturbed Schwarzschild spacetime, which is determined by solving $\hat{E}^2=V_{\rm Sch}|_{r=r_{\rm out}}$,
\begin{align}
    r_{\rm out}=\frac{r_++2\sqrt{r_+^2-3\hat{L}_0^2\left(1-\hat{E}^2\right)}\cos\left[\arg\left({\gamma}\right)/3-\pi/3\right]}{3\left(1-\hat{E}^2\right)}.\label{eq:rturn}
\end{align}
where
\begin{align}
    &\gamma=-9\hat{L}_0^2r_+\left(3\hat{E}^4-5 \hat{E}^2+2\right)-2r_+^3\nonumber\\
    &+3\sqrt{3}\hat{L}_0\left(1-\hat{E}^2\right)\\
    &\times \sqrt{4\hat{L}_0^4\left(1-\hat{E}^2\right)+\hat{L}_0^2r_+^2\left(27\hat{E}^4-36\hat{E}^2+8\right)+4r_+^4}.\nonumber
\end{align}
To the best of our knowledge, this is the first derivation of an analytic expression for the radius of the outer turning point of bound motion around the Schwarzschild BH.\footnote{One can obtain the radius of an inner turning point by switching the sign in front of $\pi/3$ in the numerator in Eq.~\eqref{eq:rturn}.} 

We now examine the sign of $dV/d\lambda|_{r=r_{\rm out}}$. A positive~(negative) value of  $dV/d\lambda|_{r=r_{\rm out}}$ indicates that the orbit is bound~(unbound). Once $\epsilon$ exceeds a certain value, $dV/d\lambda|_{r=r_{\rm out}}$ becomes negative from positive in particular directions, and in turn, outward motion is no longer reflected back, resulting in orbits that escape unboundedly from the BH. The condition,~$dV/d\lambda|_{r=r_{\rm out}}< 0$, then sets a lower bound on $\epsilon$ for unbound orbits in particular directions. The lower bound on $\epsilon$ is most stringent at $\varphi=0,\pi$, which approximately corresponds to $\epsilon_c^+$,
\begin{align}
    \epsilon_{\rm c}^+=\frac{1}{4}\frac{r_+^2\left[r_+r_{\rm out}^2-\hat{L}_0^2\left(2r_{\rm out}-3r_+\right)\right]}{4r_{\rm out}^4\left(r_{\rm out}-r_+\right)+\hat{L}_0^2r_+\left(6r_{\rm out}^2-2r_+r_{\rm out}-3r_+^2\right)}.\label{eq:Epsiloncplus}
\end{align}
The typical value is presented in the middle panel of Fig.~\ref{fig:EpsilonandL0}, showing that the typical magnitude is $\epsilon_c^+={\cal O}(10^{-7})-{\cal O}(10^{-5})$. We have verified agreement of the above estimate to a number of full numerical results obtained from Eq.~\eqref{eq:GeodesicEquation} within the correct order of magnitude.

\subsubsection{$\epsilon_c^0$: from collapsing phase to depleted phase}
The depleted phase can be interpreted as a consequence of the disappearance of stable orbits at the local minimum of $V$. 
The disappearance of the local minimum of $V$ implies that $dV/dr$ has a single zero -- corresponding to the potential peak in the vicinity of the horizon -- throughout the region of $r>r_+$ in particular directions. To quantify this behavior, we first analytically solve $d^2V_{\rm Sch}/dr^2=0$, thereby approximately determining the location of the local maximum of $dV/dr$, denoted as $r=r_{\rm inf}$,
\begin{align}
    r_{\rm inf}=\frac{3\hat{L}_0^2}{2r_+}\left[1+\frac{r_+}{3\hat{L}_0}\sqrt{9\left(\frac{\hat{L}_0}{r_+}\right)^2-24}\right].
\end{align}
We have verified that this expression agrees with the value numerically determined by $d^2V/dr^2|_{r=r_{\rm inf}}=0$ within at most $20\%$ in relative error for various values of $\hat{L}_0$. 

We now impose $dV/dr|_{r=r_{\rm inf}} < 0$, which yields a lower bound on $\epsilon$ that ensures a single zero of $dV/dr$ in particular ranges of $\varphi$. Here, the most stringent lower bound on $\epsilon$ at $\varphi=0,\pi$ approximately corresponds to $\epsilon_c^0$,
\begin{align}
\epsilon_c^0= \frac{1}{4}\frac{r_+^2\left[r_+r_{\rm inf}^2-\hat{L}_0^2\left(2r_{\rm inf}-3r_+\right)\right]}{4r_{\rm inf}^4\left(r_{\rm inf}-r_+\right)+\hat{L}_0^2r_+\left(6r_{\rm inf}^2-2r_+r_{\rm inf}-3r_+^2\right)},\label{eq:Epsilonc0}
\end{align}
which takes the same functional form as Eq.~\eqref{eq:Epsiloncplus} about the relevant radius, as both are derived from $dV/dr=0$ at certain radii. The right panel of Fig.~\ref{fig:EpsilonandL0} illustrates that the typical value is of order of $\epsilon_c^0={\cal  O}(10^{-7})-{\cal  O}(10^{-5})$. We have verified agreement of the above estimate to a number of full numerical results obtained from Eq.~\eqref{eq:GeodesicEquation} within the correct order of magnitude. 

It is worth emphasizing that the above estimate assumes an equatorial motion, i.e., $\hat{Q}=0$. For a non-equatorial system with fixed $\hat{E}$ and $\hat{Q}|_{\lambda=0}$, one can estimate $\epsilon_c^0$ by reinterpreting $\hat{L}_0$ in Eq.~\eqref{eq:Epsilonc0} as the value determined by the second equation in Eq.~\eqref{eq:rsph} with $a=0$. This is justified because the most dominant contribution for the estimate comes from the zeroth-order term; the effective potential for radial motion in the Schwarzschild spacetime takes the same form with the dependence on the total angular momentum, irrespective of the choice of the orbital plane, implying that changes of $\hat{Q}|_{\lambda=0}$ can be absorbed into a redefinition of $\hat{L}_0$ in Eq.~\eqref{eq:Epsilonc0}.

It is worth noting that $\epsilon_c^0$ is a monotonically decreasing function of $\hat{L}_0$. This property allows us to estimate the critical value of $\epsilon$ beyond which any originally bound orbits cannot exist. The minimum value of $\hat{L}_0$ for originally bound orbits is the specific angular momentum at the ISCO, i.e., $\hat{L}_0=\sqrt{3}r_+$, which serves as a useful benchmark, although particles at the ISCO would have already plunged into the BH. Substituting $\hat{L}_0=\sqrt{3}r_+$ into Eq.~\eqref{eq:Epsilonc0} yields the critical value for the depletion,
\begin{align}
\epsilon_c^{\rm depl}\simeq 8.5\times 10^{-5},\label{eq:epsilonevapo}
\end{align}
beyond which any originally bound orbits cannot exist.

\section{Astrophysical implications}\label{sec:AstrophysicalImplications}

\subsection{Structural evolution of accretion disks}
Our results in Sections~\ref{sec:geodesicsinphasespace} and~\ref{sec:GeodesisEvolution} provide a local approximate picture of the structural evolution of particle motion around an individual BH in binary systems, indicating that, as the separation decreases, accreting matter flows may experience four stages. Note that the disk we envision is thin, non self-gravitating, and non-viscous, so that its evolution is governed entirely by the BH and the tidal companion. Our results based on the idealized setup are not directly translated into robust predictions for realistic astrophysical dynamics, as realistic matter flows are influenced by additional physical effects, including surrounding magnetic fields, pressure forces, magnetized plasma effects, turbulence, and dissipative processes~\cite{1991ApJ...376..214B,1991ApJ...376..223H,1992ApJ...400..595H,1992ApJ...400..610B,1998RvMP...70....1B,Abramowicz:2011xu}.  Nevertheless, we expect that our geodesics analysis captures aspects of the underlying phase-space structure and kinematical features of matter motion around a tidally deformed BH.

During the bound phase, the tidal field primarily influences the inner matter flow. The plateaus in the fundamental-frequency ratio in Fig.~\ref{fig:FreqRatio} imply that a subset of the matter flow may undergo resonant locking near the corresponding rational ratio, leading to coherent, phase-locked dynamics. Such coherence may in turn affect the characteristic oscillations of disks, such as p-, g-, and c-modes. Moreover, the disk oscillations modulate electromagnetic~(EM) emissions, such as X-rays, from the disk~\cite{Wagoner:1998hh}, affecting quasi-periodic oscillations~(QPOs) in X-ray light curves from accreting BHs. Twin peaks of high-frequency QPOs in a ratio of $2:3$ have been reported repeatedly, motivating phenomenological models generically in the form of a non-linear resonance associated with epicyclic oscillations~\cite{Abramowicz:2001bi,Kluzniak:2001ar,Abramowicz:2011xu,Ingram:2019mna,Kluzniak:2001ar,Horak:2004hp,2009A&A...499..535H,Rebusco:2004ba},
\begin{align}
    \frac{d^2\delta_r}{dt^2}+\Omega^2_r\delta_r=&\Omega^2_r S_r\left(\delta_r,\delta_\theta,\frac{d\delta_r}{dt},\frac{d\delta_\theta}{dt}\right),\\
     \frac{d^2\delta_\theta}{dt^2}+\Omega^2_\theta\delta_\theta= &\Omega^2_\theta S_\theta\left(\delta_r,\delta_\theta,\frac{d\delta_r}{dt},\frac{d\delta_\theta}{dt}\right),
\end{align}
where $\delta_r$ and $\delta_\theta$ are radial and polar perturbations from a fiducial equatorial circular orbit, respectively; $\Omega_r$ and $\Omega_\theta$ are the corresponding radial and vertical epicyclic frequencies, respectively; $S_r$ and $S_\theta$ represent source terms responsible for non-linear resonances. Although this framework can explain the observed frequency and amplitude, the physical origin of the non-linear coupling remains unclear~\cite{Abramowicz:2011xu}. We find that tidal perturbations naturally contribute to $S_r$ and $S_\theta$, indicating that tidal effects could, in principle, affect the excitation of such a resonant pair. We emphasize, however, that the present discussion is not sufficient to quantify the role of tidal effects, particularly in the absence of comparisons with other possible mechanisms, such as forced oscillations of an accretion disk by external periodic perturbations~\cite{Lee:2004bp} or g-mode oscillations excited through coupling with a warp~\cite{2003PASJ...55..801K,Kato:2004vs}.

After the transition to chaotic phase, the inner matter flow tends to lose angular momentum due to tidal torques, thereby accreting to the BH. The plunging flow may deplete the inner region, and in turn, shift an effective inner truncation radius that is often inferred from the observed variability pattern in QPOs~\cite{Barret:2005wd,Barret:2007df}. This then raises concerns for estimates of BH spin based on an inner-disk radius identified with an ISCO radius. We note that the matter flow around the BH in LMC X-3, for example, analyzed in Ref.~\cite{2010ApJ...718L.117S} remains in bound phase with $\epsilon={\cal O}(10^{-19})$, where $M\simeq 6.98M_\odot $, $M_{\rm ext}\simeq 3.63M_\odot$, $b\simeq 9.16\times 10^9~{\rm km}$~\cite{2014ApJ...794..154O}, consistent with the observed long-term stability of the inferred inner radius.

For binaries at collapsing phase, the emergence of unbound orbits towards the tidal field can be viewed as an analogue to the Roche-lobe overflow into the companion through the inner Lagrange point,~$L_1$~\cite{10.1093/mnras/219.1.75}. It is beneficial to compare $\epsilon_c^+$ in Eq.~\eqref{eq:epsiloncminus} to a Newtonian estimate for the typical scale of the Roche lobe, beyond which the overflow occurs, determined by the Eggleton approximation~\cite{Eggleton:1983rx},
\begin{align}
    \frac{r_L}{b}=\frac{0.47 \left(M/M_{\rm ext}\right)^{2/3}}{0.6 \left(M/M_{\rm ext}\right)^{2/3}+\ln\left[1+\left(M/M_{\rm ext}\right)^{1/3}\right]}.\label{eq:EggletonApproximation}
\end{align}
Increasing $\epsilon$, we find that Eq.~\eqref{eq:EggletonApproximation} at $\epsilon=5\times 10^{-7}$ gives $r_L/r_+\simeq 20-30$. Indeed, these radii are smaller than the typical values of $r_{\rm out}$ in Eq.~\eqref{eq:rturn} with $\hat{E}=0.99$, which can approximately correspond to an outer-edge radius of the disk. Moreover, escaping orbits towards the opposite direction to the tidal field can be understood as an analogue to outflow through the outer Lagrange point,~$L_2$~\cite{1941ApJ....93..133K,1979ApJ...229..223S,1976PASJ...28..593N}. With the similar structure in the companion corresponding to the tidal field, the disks of the binary system may form spiral arms.

BHs at depleted phase may lose their stable surrounding matter flow. Indeed, such disk ``evaporation'' has been reported in Refs.~\cite{Gold:2014dta,Paschalidis:2021ntt,Bowen:2016mci}. One can heuristically characterize the typical scale associated with the depletion in terms of the Hill radius~\cite{Gold:2014dta,2019Galax...7...63G,Paschalidis:2021ntt},
\begin{align}
r_{\rm Hill}=\frac{1}{2}\left(\frac{M}{3M_{\rm ext}}\right)^{1/3}b.
\end{align}
If $r_{\rm Hill}$ is much larger than the ISCO radius, disks can persist; otherwise, they may disappear. Indeed, for $\epsilon_c^{\rm depl}$ in Eq.~\eqref{eq:epsilonevapo}, we find $r_{\rm Hill}\simeq 7.9 M$, which is comparable to the ISCO radius for a Schwarzschild BH and is modestly larger than those of Kerr BHs. The depletion occurs prior to merger, indicating that EM emissions from the binary may disappear or change their qualitative character in the spectra at a certain stage in the late inspiral.

\subsection{EMRIs in a tidal environment}
Our analysis provides insights into EMRIs in the presence of a third body. In general, bound geodesics -- not necessarily restricted to equatorial or circular orbits -- serve as a useful leading-order description for stars or compact objects orbiting a massive BH. Such systems gradually lose their energy and angular momentum due to GW emissions, eventually reaching the last stable orbit and plunging into the BH. We expect that our analysis captures an approximate instantaneous picture of the phase-space structure and kinematics underlying EMRIs in a tidal environment.

Our results in Section~\ref{sec:geodesicsinphasespace} suggest that an inspiralling object around a tidally deformed BH may experience resonances at certain separations. In particular, the plateaus in Fig.~\ref{fig:FreqRatio} suggest that the binary system in a tidal environment may remain near resonance in a long term. Consequently, the non-negligible corrections to the orbital cycle number may accumulate, potentially leaving detectable tidal imprints on GW signals. So-called tidal resonances in EMRIs have been studied by analyzing the change of action variables, induced by a tidal field, in Refs.~\cite{Bonga:2019ycj,Gupta:2021cno,Gupta:2022fbe}. It would be worthwhile to compare our findings with the results in the literature.

The last stable orbit is of particular interest for GW observations, as it marks the onset of transition from an adiabatic inspiral regime to a plunging regime~\cite{Ori:2000zn}. The shift of the radius of the last stable orbit in the presence of tidal perturbations has been studied for the non-rotating case in Refs.~\cite{Cardoso:2021qqu,Camilloni:2023rra} and for the rotating case in Ref.~\cite{cocco2026tidalperturbationsextrememass}, which may share certain similarities with the charged, non-rotating case in Ref.~\cite{Grilli:2024fds}. The existence of chaotic plunging orbits demonstrated in Section~\ref{sec:GeodesisEvolution} implies that a tidal field sourced by a third body may trigger a plunge earlier relative to the estimate for the same initial condition within the standard isolated two-body system. It is also worth noting that the notion commonly referred to as the ``ISCO'' may cease to be appropriate once the tidal perturbation becomes non-negligible. This is because (i)~circular orbits are not generically well defined in the absence of conserved angular momenta, and (ii) even if one adopts a conventional criterion based on the potential analysis, the potential for $\epsilon\ge \epsilon_c^0$ cannot satisfy the usual ISCO conditions -- simultaneous vanishing of its first- and second-order derivatives -- at any radii.

Unbound orbits demonstrated in Section~\ref{sec:GeodesisEvolution} suggest that a fraction of initially bound objects may escape from the BH at certain stages of the inspiral, instead of plunging into the BH. In such cases, the GW signal deviates from the isolated EMRI scenario and could instead exhibit non-standard features such as non-quasinormal decays arising from unbound departures. Note that, as estimated in the next section, LISA could capture all the stages of particle dynamics.

\subsection{Imprints on GW observations}
We discuss whether the BH-particle interactions discussed in the previous sections can leave an observable imprint on GW observations. Within our framework, the frequency of the GW from the binary system can be expressed in terms of $\epsilon$ as
\begin{align}
    f_{\rm GW}=& N_{\rm GW}\left(\frac{M_\odot}{M}\right)\left(\frac{M+M_{\rm ext}}{M_{\rm ext}}\right)^{1/2}\epsilon^{1/2},\label{eq:fGW}
\end{align}
where $N_{\rm GW}:=c^3/({GM_\odot})(\simeq 6.46\times 10^4~{\rm s}^{-1})$ with the solar mass~$M_\odot\simeq 1.99\times 10^{30}$~kg. Now, we are interested in $0<\epsilon \le 10^{-5}$, which is sufficient to consider accreting BHs prior to depleted phase.

For the LIGO-Virgo-KAGRA~(LVK) observation~\cite{LIGOScientific:2014pky,VIRGO:2014yos,Somiya:2011np}, the frequency band with high sensitivity is typically characterized as $10~{\rm Hz}\lesssim f_{\rm GW}\lesssim 1000~{\rm Hz}$. For a  binary system with $M=M_{\rm ext}=30M_\odot$, Eq.~\eqref{eq:fGW} yields $f_{\rm GW}\simeq 9.6$~Hz when $\epsilon=10^{-5}$, which is very close to the low-frequency edge relevant for the LVK observation. Consequently, any corresponding environmental effect is expected to be challenging to detect with the current ground-based detectors. Nevertheless, if accreting matter flow produces EM emission at the earlier stages in gas-rich environments, e.g., active galactic nuclei~\cite{Grobner:2020drr} or circumbinary disks~\cite{Schnittman:2010wy}, such radiation could carry information on the source localization and potentially informative priors on the intrinsic parameters, such as the BH mass and spin, prior to the LVK observation, thereby offering a potential way for improved parameter estimation. It is worth noting that the observational bandwidth of (B-)DECIGO is anticipated as $0.1-10$~Hz~\cite{Kawamura:2020pcg}, which may correspond to $10^{-9}\lesssim\epsilon \lesssim 10^{-5}$. Moreover, Einstein Telescope~\cite{Punturo:2010zz,Sathyaprakash:2012jk,ET:2025xjr} and Cosmic Explorer~\cite{Reitze:2019iox} will extend the sensitivity to ${\cal O}(1)~{\rm Hz}\le f_{\rm GW}\le {\cal O}(10^3)~{\rm Hz}$, potentially reaching $\epsilon={\cal O}(10^{-5})$.

The frequency band of LISA is typically expected as $10^{-4}-{10^{-1}}~{\rm Hz}$~\cite{2017arXiv170200786A}. For a binary system with $(M,M_{\rm ext})=(30M_\odot,10^5M_\odot)$ and $10^{-10}\le \epsilon \le 10^{-5}$, the GW frequency varies within $3.7\times 10^{-4}~{\rm Hz}\lesssim f_{\rm GW}\lesssim 1.2\times 10^{-1}~{\rm Hz}$. This indicates that LISA may probe all the stages of the evolution of matter flows around or objects orbiting tidally deformed BHs. Moreover, for a system with $(M,M_{\rm ext})=(10^5 M_\odot,10^5 M_\odot)$ and $10^{-7}\le \epsilon \le 10^{-5}$, the GW frequency varies within $2.9\times 10^{-4}~{\rm Hz}\lesssim f_{\rm GW} \lesssim 2.9\times 10^{-3}~{\rm Hz}$. Accordingly, components of the host galaxies, i.e., stars, gas, as well as compact objects, may experience chaotic dynamics and depletion within the LISA band, potentially giving rise to measurable signature in GWs and EM emission~\cite{Dotti:2025sip}.

\section{Summary}\label{sec:Summary}
We have investigated the geometry of a tidally deformed rotating BH, and the geodesic motion of particles in its vicinity. To this end, we have rederived a tidally perturbed Kerr BH solution. Our derivation relies on two different approaches using the Geroch, Held, and Penrose operator~\cite{Geroch:1973am,Berens:2024czo} and the Chrzanowski, Cohen, and Kegeles operator~\cite{Cohen:1974cm,Chrzanowski:1975wv,PhysRevD.19.1641,Wald:1978vm}, yielding consistent results with previous work~\cite{LeTiec:2020bos}. We provide the generator of the solutions online~\cite{Notebook,CoG} and the resulting expressions in Appendix~\ref{Sec:metricreconstruction}.
The deformed geometry models an astrophysical BH in adiabatically evolving binary systems at large separations, allowing us to study relativistic restricted three-body problems motivated by various astrophysical scenarios, including accretion-disk dynamics and EMRIs. For example, the derived spacetime may be useful for further analysis of accretion disks around a BH in binary systems, based on GR magnetohydrodynamic simulations.

One of the key parameters is the tidal strength parametrized by a dimensionless, non-negative parameter,~$\epsilon$, that encapsulates the binary mass ratio and separation. In the presence of tidal perturbations, chaotic geodesic motion arises from the non-integrability of the system. For sufficiently small values of $\epsilon$, the phase-space structure retains entirely regularity, despite slight deformations, in accordance with the KAM theorem. In addition, the ratio of the radial, polar, and azimuthal fundamental frequencies of geodesics exhibits plateaus with respect to the initial radius at nearly rational numbers. This result suggests that (i)~a fraction of the matter flow around a tidally deformed BH may undergo resonant locking near the corresponding rational ratio, leading to coherent, phase-locked dynamics, and (ii)~EMRIs in a tidal environment may remain near resonances in a long term, modifying the long-term orbital evolution. As $\epsilon$ increases, the chaotic feature is more pronounced and eventually becomes dominant in the dynamics.

The structural evolution of initially bound particle motion around a tidally deformed BH is classified into four stages depending on the magnitude of $\epsilon$: bound phase~$(0\le \epsilon< \epsilon_c^-$), chaotic phase~($\epsilon_c^-\le\epsilon< \epsilon_c^+ $), collapsing phase~($\epsilon_c^+\le \epsilon<\epsilon_c^0$), and depleted phase~($\epsilon\ge \epsilon_c^0$), where the critical tidal amplitudes, $\epsilon_c^-$, $\epsilon_c^+$, and $\epsilon_c^0 $, are semi-analytically estimated. During bound phase, all trajectories remain bound with weak chaos, and in turn, at chaotic phase, a fraction of the trajectories plunges into the BH. After the transition to collapsing phase, a subset of the remaining trajectories becomes unbound and escapes to and against the directions of the tidal field. At depleted phase, no bound orbits persist, leaving plunging and unbound orbits. We have verified the consistencies of these critical tidal amplitudes with well-known characteristic scales associated with the Roche-lobe overflow and disk depletion within the Newtonian framework. Our estimates suggest that current and future ground-based GW detectors may only cover depleted phase, indicating that matter flow around BHs in binary systems may already be depleted in the corresponding frequency band. By contrast, LISA and (B-)DECIGO may probe the earlier stages, potentially opening multi-messenger windows combined with EM counterparts.

Within the non-magnetized, test-particle approximation, our analysis provides insights into the idealized structure that may be relevant to accretion matter flows around a BH in binary systems. Chaotic structure may modulate characteristic disk oscillations, and in turn, affect EM emissions. Indeed, tidal perturbations naturally modulate nonlinear couplings among epicyclic oscillations, and could therefore, in principle, affect the excitation of a resonant pair in high-frequency QPOs. We emphasize, however, that the present discussion is not sufficient to quantify the role of tidal effects, particularly in the absence of comparisons with other possible mechanisms, such as forced oscillations of an accretion disk by external periodic perturbations~\cite{Lee:2004bp} or g-mode oscillations excited thorough coupling with a warp~\cite{2003PASJ...55..801K,Kato:2004vs}. It would be worthwhile to study this potential contribution within GR magnetohydrodynamic settings.

The present work opens an avenue to study EMRIs in a tidal environment. Trapping of an orbiting object near resonances in a long term, suggested by our phase-space analysis, may lead to accumulated, non-negligible corrections to the GW phase. It would be worthwhile to compare our findings with tidal resonances in EMRIs studied in Refs.~\cite{Bonga:2019ycj,Gupta:2021cno,Gupta:2022fbe}. Moreover, tidal perturbations may trigger a plunge earlier relative to the isolated case with the same initial condition, and may yield escaping orbits from the BH. In light of detectors planned for the coming decades, it will be necessary to incorporate these effects into theoretical modeling of GWs in the late inspiral of EMRIs. Modeling accurate waveforms will require comprehensive analyses of geodesic dynamics in tidally perturbed BH spacetimes.

\acknowledgments
%

We thank Hiroyuki Nakano for valuable discussions on the implications for EMRIs, Paolo Pani for useful feedback on the manuscript, Hiromichi Tagawa for insightful comments on the implications for disk dynamics, Eric Poisson for careful feedback on the manuscript, and Kyriakos Destounis for his helpful technical guidance on the phase-space analysis and for providing us his Mathematica notebook, which was used for the double check of computations of the rotation number.
T.K.\ is supported by the MUR FIS2 Advanced Grant ET-NOW (CUP:~B53C25001080001) and by the INFN TEONGRAV initiative.
The Center of Gravity is a Center of Excellence funded by the Danish National Research Foundation under grant No. DNRF184.
V.C.\ is a Villum Investigator and acknowledges support by VILLUM Foundation (grant no. VIL37766). 
V.C. acknowledges financial support provided under the European Union’s H2020 ERC Advanced Grant “Black holes: gravitational engines of discovery” grant agreement no. Gravitas–101052587. 
Views and opinions expressed are however those of the author only and do not necessarily reflect those of the European Union or the European Research Council. Neither the European Union nor the granting authority can be held responsible for them.
This project has received funding from the European Union's Horizon 2020 research and innovation programme under the Marie Sklodowska-Curie grant agreement No 101007855 and No 101131233.
This work is supported by Simons Foundation International \cite{sfi} and the Simons Foundation \cite{sf} through Simons Foundation grant SFI-MPS-BH-00012593-11.

\appendix

\section{Tidal perturbations on a Kerr geometry}\label{Sec:tidalperturbation}

\subsection{Background null tetrad}
We analyze tidal perturbations within the Newman-Penrose formalism~\cite{Newman:1961qr,Chandrasekhar:1985kt}. First, let us introduce two real null vectors, $\{{\bf l},{\bf n}\}$, and two complex null vectors, $\{{\bf m},\overline{\bf m}\}$, where $\overline{\bf m}$ is the complex conjugate of ${\bf m}$, subject to~\cite{Newman:1961qr}
\begin{align}
    l_\mu l^\mu=n_\mu n^\mu&=m_\mu m^\mu=\overline{m}_\mu \overline{m}^\mu=0,\nonumber\\
    l_\mu n^\mu&=-1,~~m_\mu \overline{m}^\mu=1,\\
    l_\mu m^\mu=l_\mu \overline{m}^\mu &=n_\mu m^\mu =n_\mu \overline{m}^\mu=0.\nonumber
\end{align} 
The set of these vector fields on the Kerr spacetime forms the Kinnersley tetrad whose components are explicitly~\cite{Kinnersley:1969zza} 
\begin{equation}
\begin{split}
\label{eq:tetrad}
    l^\mu=&\left(2\frac{r^2+a^2}{\Delta},1,0,\frac{2a}{\Delta}\right),\quad
    n^\mu= \left(0,-\frac{\Delta}{2\Sigma},0,0\right),\\
    m^\mu=&\frac{1}{\sqrt{2}\left(r+i a \cos\theta\right)}\left(ia\sin \theta,0,1,\frac{i}{\sin \theta}\right),\\
    \overline{m}^\mu=&\frac{1}{\sqrt{2}\left(r-i a \cos\theta\right)}\left(-ia\sin \theta,0,1,-\frac{i}{\sin \theta}\right),
    \end{split}
\end{equation}
where $\Delta:=(r-r_+)(r-r_-)$.

\subsection{Tidal moments}\label{Sec:TidallyDeformedbody}

We introduce the notion of gravitoelectric tidal moments, which encode information of an external gravitational source approximately.  In a binary system with large separation, the characteristic length scale of the external universe,~${\cal R}$, is much larger than $M$, i.e., $M \ll {\cal R}$.  

In the region of $r\ll {\cal R}$, tidal deformations are described within fully relativistic perturbation theory. The multipole structure of a tidally deformed BH is determined by solving the linearized Einstein equations, while the details of the external gravitational source remain unspecified.  Given an external tidal field, the tidal moment is introduced, assuming that the deformed BH can be described by a point mass in a weakly curved spacetime in the region of $r\gg M$. The gravitoelectric tidal moments are defined from the frame components of the Weyl tensor,~$C_{\mu \alpha \nu \beta}$, for metric expanded around the worldline of the point mass along a timelike geodesic parametrized by~$t$~\cite{PhysRevD.31.1815,Poisson:2004cw}:
\begin{align}
    {\cal E}_{ij}\left(t\right) :=&\left[C_{0i0j}\right]^{\rm STF},\label{eq:Eij}
\end{align}
where the superscript, STF, indicates that they are a symmetric and tracefree~(STF) tensor associated with the local frame.

The tidal moment is tied to an integration constant of tidal perturbations~\cite{Poisson:2004cw,Poisson:2009qj}. To see the connection, we now analyze the Weyl scalar,
\begin{align}
\label{eq:Psi4}
\psi_{0}:=C_{\alpha\beta\delta\gamma}l^\alpha m^\beta l^\delta m^\gamma.
\end{align}
It is worth noting that the above definition coincides with the Weyl scalar computed in Ref.~\cite{LeTiec:2020bos}, but differs in sign from those in Refs.~\cite{Poisson:2004cw,Chatziioannou:2012gq}. 

The expansion of $\psi_{0}$ on the perturbed Kerr spacetime in $r\gg M$ generically takes the form~\cite{Poisson:2004cw,Poisson:2009qj,LeTiec:2020bos},
\begin{align}
\psi_{0}=\sum_{\ell,m} \alpha_{\ell m} r^{\ell-2}\left[1+{\cal O}\left(M/r\right)\right]Y_{\ell m}^{+2}\left(\theta,\varphi\right),\label{eq:psiasyinNewton}
\end{align}
where a spin-weighted~$(s=+2)$ spherical harmonic, $Y_{\ell m}^{+2}\left(\theta,\varphi\right)=[(\ell+2)(\ell+1)\ell(\ell-1)]^{-1/2}\eth_1 \eth_0 Y_{\ell m}$, with the spin-raising operator~$\eth_s:=-\partial_\theta-(i/\sin\theta)\partial_\varphi+s\cot\theta$ for scalar spherical harmonics~$Y_{\ell m}(\theta,\varphi)$~\cite{Berti:2005eb,Goldberg:1966uu}.
We have introduced a complex-valued integration constant, 
\begin{align}
\label{eq:alphaellm}
\alpha_{\ell m}:=&\sqrt{\frac{\left(\ell+2\right)\left(\ell+1\right)}{\ell\left(\ell-1\right)}}{\cal E}_{\ell m}.
\end{align}
Here, ${\cal E}_{\ell m}={\cal E}_{\ell m}(t)$ is the component of the tidal moments with decomposition in terms of spherical harmonics.\footnote{The timescale of the tidal perturbation is assumed to be much longer than $M$, thus enabling decomposition in terms of spherical harmonics, instead of spheroidal harmonics. } Note that $\alpha_{\ell m}$ satisfies $\bar{\alpha}_{\ell m}=(-1)^m \alpha_{\ell -m}$. For $\ell=2$, ${\cal E}_{2 m}$ is related to ${\cal E}_{ij}$ in Eq.~\eqref{eq:Eij} via~\cite{Poisson:2004cw,Poisson:2009qj}
\begin{align}
    {\cal E}_{ij}n^i n^j={\cal E}_{2m}Y_{2m},
\end{align}
 where $n^i=(\sin\theta \cos\varphi,\sin\theta \sin\varphi,\cos\theta)$. The coefficients,~${\cal E}_{2 m}$, can be expressed in terms of ${\cal E}_{ij}$ as~\cite{LeTiec:2020bos,Poisson:2004cw,Chatziioannou:2012gq}
 \begin{align}
\alpha_{20}=&-2\sqrt{\frac{6\pi}{5}}\left({\cal E}_{11}+{\cal E}_{22}\right),\label{eq:alpha20}\\
\alpha_{2\pm1}=&\mp2\sqrt{\frac{4\pi}{5}}\left({\cal E}_{13}\mp i {\cal E}_{23}\right),\\
\alpha_{2\pm2}=&\sqrt{\frac{4\pi}{5}}\left({\cal E}_{11}-{\cal E}_{22}\mp 2 i {\cal E}_{12}\right).
 \end{align}
These expressions are consistent with Eqs.~(4.8) and~(4.9) of Ref.~\cite{LeTiec:2020bos}. As noted there in footnote~11, a difference in normalization of spherical harmonics leads to different numerical coefficients in the above expressions from Eqs.~(9.17)-(9.22) of Ref.~\cite{Poisson:2004cw}; Eqs.~(7.23) and~(7.24) of Ref.~\cite{Taylor:2008xy}; Tables~VIII and IX of Ref.~\cite{Poisson:2009qj}; Eq.~(28) of Ref.~\cite{Chatziioannou:2012gq}. The spin-weighted spherical harmonics employed in this work are normalized following the manner in Appendix~A of Ref.~\cite{LeTiec:2020bos}. Within the setup given by Eqs.~\eqref{eq:E11pE22}-\eqref{eq:E12} with $\epsilon$ defined in Eq.~\eqref{eq:epsilon}, we obtain
\begin{align}
{\cal E}_{2m}Y_{2m}\simeq&\frac{\epsilon}{2M^2}\left[-1+3\cos^2\theta\right.\nonumber\\
&\left.-3\cos\left(2\sqrt{\frac{M}{b^3}}t-2\varphi\right)\sin^2\theta\right],\label{eq:TidalStrength}
\end{align}
up to linear order of $\epsilon$.

\subsection{Stationary tidal perturbations}
Tidal perturbations are governed by the Teukolsky equations~\cite{Teukolsky:1973ha,Teukolsky:1974yv}. The solution with  $s=+2$ is decomposed as
\begin{align}
    \psi_0=\sum_{\ell m }\alpha_{\ell m} R_{\ell m}^{+2}(r)~ Y_{\ell m }^{+2}.\label{eq:decompositionofpsi0}
\end{align}
Note that, since our interest is in a slowly-varying tidal perturbation, the spin-weighted spheroidal harmonics reduce to $Y_{\ell m}^{+2}$. Following the manner of Ref.~\cite{LeTiec:2020bos}, we obtain $R_{\ell m}^{+2}$ regular at the event horizon, thereby deriving
\begin{widetext}
\begin{align}
    \psi_0=\sum_{\ell m}\alpha_{\ell m}\frac{\Gamma\left(\ell-1\right)\Gamma\left(\ell+1+2i m a/(r_+-r_-)\right)}{\Gamma\left(2\ell+1\right)\Gamma\left(-1+2i m a/(r_+-r_-)\right)} \frac{\left(r_+-r_-\right)^{\ell+2}}{\left(r-r_+\right)^2\left(r-r_-\right)^2} ~_2F_1\left(-\ell-2,\ell-1;-1+ \frac{2ima}{r_+-r_-};-\frac{r-r_+}{r_+-r_-}\right) Y_{\ell m}^{+2},\label{eq:tidalpsi0}
\end{align}
\end{widetext}
which takes asymptotically the form of Eq.~\eqref{eq:psiasyinNewton} at $r\gg M$. The radial part is consistent with Eq.~(9.28) of Ref.~\cite{Poisson:2004cw} and Eq.~(17) of Ref.~\cite{Yunes:2005ve} with an opposite sign arising from the different convention in decomposition of $\psi_0$ in Eq.~\eqref{eq:decompositionofpsi0}.

\section{Reconstruction of a tidally deformed metric}\label{Sec:metricreconstruction}

\subsection{Hertz potential}
Reconstruction of a tidally deformed metric consists of two parts:~(i) deriving a potential -- referred to as a Hertz potential -- from the Weyl scalar and (ii) acting a certain differential operator on the potential, thereby obtaining a metric perturbation that shares physically same information with the Weyl scalar used in the reconstruction~\cite{Cohen:1974cm,Chrzanowski:1975wv,Wald:1978vm,Stewart:1978tm,Whiting:2005hr}. Here, we first introduce the Hertz potential~$\Psi$ for $\psi_0$, which is a solution to a source-free Teukolsky equation with $s=-2$ and satisfies the following fourth-order differential equation,
\begin{align}
    {\bm D}^4 \bar{\Psi} =2\psi_0,\label{eq:eqforPsi}
\end{align}
where ${\bm D}=l^\alpha \partial_\alpha$. The Weyl scalar~$\psi_0$ is given in Eq.~\eqref{eq:tidalpsi0}. It is worth noting that the factor of $2$ on the right-hand side is consistent with Eq.~(5.1) of Ref.~\cite{LeTiec:2020bos}, while not with Eq.~(11) in Ref.~\cite{Ori:2002uv}. As previously pointed out by Ref.~\cite{Keidl:2010pm}, missing the factor of $2$ in $\Psi$ may lead to incorrect results without great care, which are shared by Refs.~\cite{Lousto:2002em,Ori:2002uv,Yunes:2005ve} inherited from Ref.~\cite{PhysRevD.19.1641}~(see also footnote~2 of Ref.~\cite{Keidl:2010pm} and footnote~17 of Ref.~\cite{LeTiec:2020bos}).

Following Refs.~\cite{Ori:2002uv,Yunes:2005ve}, one can invert Eq.~\eqref{eq:eqforPsi} into
\begin{align}
    \bar{\Psi}=\frac{2}{[(\ell+2)(\ell+1)\ell(\ell-1)]^2}\Delta^2 \left({\bm D}^\dagger\right)^4 \Delta^2 \psi_0,
\end{align}
where ${\bm D}^\dagger:= \partial_r$. Note again the difference of the factor of $2$ from Eq.~(21) in Ref.~\cite{Ori:2002uv}. With Eq.~\eqref{eq:tidalpsi0}, we obtain
\begin{widetext}
\begin{align}
    \Psi=&\sum_{\ell m}\bar{\alpha}_{\ell m} \left[ \frac{2}{\left[\left(\ell+2\right)\left(\ell+1\right)\ell\left(\ell-1\right)\right]^2} \frac{\Gamma\left(\ell+3\right)^2\Gamma\left(\ell+1-2i m a/(r_+-r_-)\right)}{\Gamma\left(\ell-1\right)\Gamma\left(2\ell+1\right)\Gamma\left(3-2im a/(r_+-r_-)\right)} \frac{\left(r-r_+\right)^2 \left(r-r_-\right)^2}{\left(r_+-r_-\right)^{-\ell+2}}\right. \nonumber\\
    &\left.\quad\quad \times ~_2F_1\left(-\ell+2,\ell+3;3-\frac{2i ma}{r_+-r_-};-\frac{r-r_+}{r_+-r_-}\right)\bar{Y}_{\ell m}^{+2}\right].\label{eq:HetrzPotential}
\end{align}
\end{widetext}
This coincides with the complex conjugate of Eq.~(5.2) with Eq.~(5.5) of Ref.~\cite{LeTiec:2020bos}. We have verified that $\Psi$ in Eq.~\eqref{eq:HetrzPotential} indeed satisfies Eq.~\eqref{eq:eqforPsi}. 
For $\ell=2$, Eq.~\eqref{eq:HetrzPotential} is different by a factor of $-2$ from Eq.~(34) of Ref.~\cite{Yunes:2005ve}, which arises from~(i) the opposite sign in decomposition in Eq.~(8) of the reference to Eq.~\eqref{eq:decompositionofpsi0} of the current work and (ii)~missing the factor of $2$ on the right-hand side in Eq.~(20) in the reference as previously noted. 

\subsection{Metric reconstruction}
Before introducing the operator generating a metric perturbation from $\Psi$, we adopt the ingoing-radiation gauge on the metric perturbation~$h_{\mu\nu}$~\cite{Chrzanowski:1975wv,PhysRevD.19.1641}
\begin{align}
    l^\mu h_{\mu\nu}=0,\quad g^{\mu\nu} h_{\mu\nu}=0.\label{eq:IRG}
\end{align}
The Geroch, Held, and Penrose~(GHP) operator, ${\cal S}_{\mu\nu}$, allows us to generate a metric perturbation from the Hertz potential~\eqref{eq:HetrzPotential} via~\cite{Whiting:2005hr}
\begin{align}
    h_{\mu\nu}=4{\rm Re}\left[{\cal S}_{\mu\nu}^\dagger \Psi\right].
\end{align}
Note that the factor of~$4$ is twice the corresponding expression in Eq.~(1.18) of Ref.~\cite{Berens:2024czo}. This difference arises because the Hertz potential used in the reference, satisfying Eq.~(1.22a) therein, is defined to be twice $\Psi$ introduced in Eq.~\eqref{eq:eqforPsi}. The explicit form of ${\cal S}_{\mu \nu}^\dagger$ on the Kerr spacetime reads~\cite{Geroch:1973am,Berens:2024czo}
\begin{align}
    {\cal S}_{\mu \nu}^\dagger:=&-\frac{1}{4\bar{\zeta}^2}l_\mu l_\nu \left(\bar{\cal L}_1 -\frac{2ia \sin\theta}{\zeta}
 \right) \bar{\cal L}_2\nonumber \\
 &-\frac{1}{2}m_\mu m_\nu \left(l^\alpha \partial_\alpha-\frac{2}{\zeta}\right) l^\beta \partial_\beta\\
 &+\frac{1}{\sqrt{2}\bar{\zeta}} l_{(\mu} m_{\nu)}\left(l^\alpha \partial_\alpha \bar{\cal L}_2+\frac{a^2\sin2\theta}{\Sigma} l^\alpha \partial_\alpha -\frac{2r}{\Sigma}\bar{\cal L}_2 \right)\nonumber,
\end{align}
where $\zeta:= r-ia \cos\theta$ and
$
    {\cal L}_n:=\partial_\theta +{m}/{\sin\theta}+n \cot \theta.
$
As a cross-check, one can reconstruct $h_{\mu\nu}$ by acting a differential operator employed by Chrzanowski, Cohen, and Kegeles ~\cite{Cohen:1974cm,Chrzanowski:1975wv,PhysRevD.19.1641,Wald:1978vm,Lousto:2002em,Keidl:2006wk}. In this manner, we have obtained exactly the same result as in the above approach based on the GHP operator. The components of $h_{\mu\nu}$ read~\cite{LeTiec:2020bos}

\begin{widetext}
    \begin{align}
        h_{vv}=&h_{(n)(n)}+2\sqrt{2}a \sin \theta ~{\rm Im}\left[\frac{h_{(n)(\bar{m})}}{\bar{\zeta}}\right]-a^2 \sin^2\theta~{\rm Re}\left[\frac{h_{(\bar{m})(\bar{m})}}{\bar{\zeta}^2} \right],\\
        h_{vr}=&-\frac{2\Sigma}{\Delta}\left(h_{(n)(n)}+\sqrt{2}a \sin \theta~{\rm Im}\left[\frac{h_{(n)(\bar{m})}}{\bar{\zeta}}\right]\right),\\
        h_{v\theta}=&\sqrt{2}\Sigma~{\rm Re}\left[\frac{h_{(n)(\bar{m})}}{\bar{\zeta}}\right]+a \Sigma \sin\theta ~{\rm Im}\left[\frac{h_{(\bar{m})(\bar{m})}}{\bar{\zeta}^2}\right],\\
        h_{v\varphi}=&-\sqrt{2} \sin \theta \left(\Sigma+2a^2 \sin^2\theta\right)~{\rm Im}\left[\frac{h_{(n)(\bar{m})}}{\bar{\zeta}}\right]-a \sin^2\theta~\left\{h_{(n)(n)}-\left(r^2+a^2\right)~{\rm Re}\left[\frac{h_{(\bar{m})(\bar{m})}}{\bar{\zeta}^2}\right]\right\},\\
        h_{rr}=&\frac{4\Sigma^2}{\Delta^2}h_{(n)(n)},\\
        h_{r\theta}=&-\frac{2\sqrt{2}\Sigma^2}{\Delta}~{\rm Re}\left[\frac{h_{(n)(\bar{m})}}{\bar{\zeta}}\right], \\
        h_{r\varphi}=&\frac{2\Sigma}{\Delta}a\sin^2\theta h_{(n)(n)}+\frac{2\sqrt{2}\Sigma}{\Delta}\left(r^2+a^2\right) \sin\theta~{\rm Im}\left[\frac{h_{(n)(\bar{m})}}{\bar{\zeta}}\right],\\
        h_{\theta\theta}=& \Sigma^2~{\rm Re}\left[\frac{h_{(\bar{m})(\bar{m})}}{\bar{\zeta}^2} \right],\\
        h_{\theta\varphi}=&-\sqrt{2}a \Sigma \sin^2\theta~{\rm Re}\left[\frac{h_{(n)(\bar{m})}}{\bar{\zeta}}\right]-\Sigma \sin\theta \left(r^2+a^2\right)~{\rm Im}\left[\frac{h_{(\bar{m})(\bar{m})}}{\bar{\zeta}^2}\right],\\
        h_{\varphi\varphi}=&a^2\sin^4\theta h_{(n)(n)}+2\sqrt{2}a \sin^3\theta \left(r^2+a^2\right)~{\rm Im}\left[\frac{h_{(n)(\bar{m})}}{\bar{\zeta}}\right]-\left(r^2+a^2\right)^2\sin^2\theta~{\rm Re}\left[\frac{h_{(\bar{m})(\bar{m})}}{\bar{\zeta}^2}\right],
    \end{align}
with 
\begin{align}
    h_{(n)(n)}=& -\left({\bm \delta} +2\beta +\bar{\pi}-\gamma\right)\left({\bm \delta}+4\beta +3\gamma\right)\Psi+{\rm c.c.},\nonumber\\
    h_{(n)(\bar{m})}=& =-\frac{1}{2}\left\{\left({\bm \delta}+4\beta-2\bar{\pi}-\gamma\right)\left({\bm D}-\frac{3}{\zeta}\right)+\left({\bm D}-\frac{1}{\bar{\zeta}}+\frac{1}{\zeta}\right)\left({\bm \delta}+4\beta+3\gamma\right)\right\}\Psi,\\
    h_{(\bar m)(\bar{m})}=& -\left({\bm D}+\frac{1}{\zeta}\right)\left({\bm D}-\frac{3}{\zeta}\right)\Psi.\nonumber
\end{align}
\end{widetext}
Here, we have introduced ${\bm \delta}:= m^\mu \partial_\mu$, $\beta:= \cot\theta/(2\sqrt{2}\bar{\zeta})$, $\pi :=i a \sin\theta/(\sqrt{2}\zeta^2)$,~$\tau:=-ia \sin\theta/(\sqrt{2}\zeta \bar{\zeta})$. On the equatorial plane~$\theta=\pi/2$, the above expressions reduce to
\begin{widetext}
\begin{align}
    h_{vv}=&\frac{M_{\rm ext}}{2b^3 r^3}\left[r^3\left(r-2M\right)^2+a^2 r \left(3r^2-4M r -2M^2\right)+2 M a^4\right.\nonumber\\
    &+\left\{3r^3\left(r-2M\right)^2-a^2 r \left(3r^2-4M r-2M^2\right)\right\}\cos\tilde{\varphi}\\
    &\left.+2a r\left\{a^2\left(2r+M\right)+r\left(3r^2-8M r +\sqrt{\sigma_{v,+}}+\sqrt{\sigma_{v,-}}\right)\right\}\sin\tilde{\varphi}\right],\nonumber\\
    h_{vr}=&-\frac{M_{\rm ext}}{b^3}\left[\Delta+\left(3r^2-6M r-a^2\right)\cos\tilde{\varphi}+4a \left(r-M\right)\sin\tilde{\varphi}\right],\\
    h_{v\varphi}=&-\frac{M_{{\rm ext}}}{4 b^{3} r^{3}}\left[2 a^{3} r \left(3 r^{2}-2 M r-2 M^{2}\right)+ a r^{3}\left\{4 r (r-2 M)+\sqrt{\sigma_{v,+}}+\sqrt{\sigma_{v,-}}\right\}+ 4 a^{5} M\right.\nonumber\\
    &- 4 a\left\{a^{4} M+a^{2} r (r^{2}-M r-M^{2})
+ r^{3}(r^{2}+2 M r-7 M^{2})\right\}\cos\tilde{\varphi}\\
&\left.+ 4 r \left\{r^{4}(r-2 M)+a^{2} r (r^{2}-5 M r+4 M^{2})+ a^{4} M\right\}\sin\tilde{\varphi}\right],\nonumber\\
    h_{rr}=&\frac{2M_{\rm ext} r^2}{b^3}\left(1+3\cos\tilde{\varphi}+\frac{2a}{r}\sin\tilde{\varphi}\right),\\
    h_{r\varphi}=&\frac{M_{\rm ext}}{b^3}\left[a\Delta-a\left\{a^2+r\left(r+6M\right)\right\}\cos\tilde{\varphi}+2\left\{r^3+a^2\left(r-2M\right)\right\}\sin\tilde{\varphi}\right],\\
    h_{\theta\theta}=&-\frac{M_{{\rm ext}}r}{2b^{3}}
\left[
r^{3}-2 M^{2} r-a^{2}(r-2 M)\right.\nonumber\\
&\left.-\Big\{r^{3}-2 M^{2} r-a^{2}(5 r-2 M)\Big\}\cos\tilde{\varphi}
+ a\Big\{2 a^{2}+\big(\sqrt{\sigma_{A,+}}+\sqrt{\sigma_{A,-}}-4 r\big)r\Big\}\sin\tilde{\varphi}
\right]
,\\
h_{\varphi\varphi}=&\frac{M_{{\rm ext}}}{2 b^{3} r^{3}}
\left[
r^{7}+2 a^{6} M-2 M^{2} r^{5}
+2 a^{2} r^{3}\left(r^{2}-M r-8 M^{2}\right)
+a^{4} r\left(r^{2}-2 M^{2}\right)
+16 a^{2} M^{2} r^{3}\right.\nonumber\\
&-\Big\{
r^{7}+2 a^{6} M-2 M^{2} r^{5}
+2 a^{2} r^{3}\left(r^{2}-M r-8 M^{2}\right)
+a^{4} r\left(r^{2}-2 M^{2}\right)
\Big\}\cos\tilde{\varphi}\\
&\left.-a r\Big\{
3\big(\sqrt{\sigma_{A,+}}+\sqrt{\sigma_{A,-}}\big) r^{4}
+4 a^{2} M r (r-2 M)
-2 a^{4} M
\Big\}\sin\tilde{\varphi}
\right]\nonumber
,\\
    h_{v\theta}=&h_{r\theta}=h_{\theta\varphi}=0,
\end{align}
\end{widetext}
where we have defined 
\begin{align}
\tilde{\varphi}:=&2(t\sqrt{(M+M_{\rm ext})/b^3}-\varphi),\\
\sigma_{v,\pm}:=&a^4+8M^3 r_\mp+ 4 M a^2\left(r_\pm-3M\right),\\
\sigma_{A,\pm}:=&2M r_\pm-a^2.
\end{align}
Note that, although the time variation of the deformed geometry is negligible within static tides, $t$ enters only in $\tilde{\varphi}$. This is also the case for non-equatorial configurations because $t$ is introduced only through ${\cal E}_{2m}$ as implied in Eq.~\eqref{eq:TidalStrength}. In other words, the choice of the reference initial time for any dynamics on this background is equivalent to a constant shift of the initial azimuthal angle~$\varphi$. Hence, one can set arbitrary values to $t$ as a reference initial time for an instantaneous snapshot without loss of generality. 

It is worth comparing the metric re-derived in this work with the result of Ref.~\cite{Yunes:2005ve}. First, the metric obtained in the literature does not contain the $m=0$ mode. Second, there is difference in an overall factor of $2$ arising from the missing factor in the Hertz potential defined in Eq.~(20) of the reference, as noted before.

\section{Tidally deformed Schwarzschild metric}\label{sec:DeformedSchwarzschild}
Here, we provide a tidally deformed Schwarzschild metric in the Regge-Wheeler gauge~\cite{PhysRev.108.1063}. The metric takes the form of Eq.~\eqref{eq:metric}. With the gauge transformation from the ingoing-radiation gauge~\eqref{eq:IRG} and the coordinate transformation to the static coordinates~$(t,r,\theta,\varphi)$, the components of $h_{\mu \nu}$ read
\begin{widetext}
\begin{align}
    h_{\mu\nu}=\begin{pmatrix}
-r^2\left(1-\dfrac{2M}{r}\right)^2&0 & 0&0\\
0&- r^2&& 0\\
0 &0&- r^4 \left[1-2\left(\dfrac{M}{r}\right)^2\right] &0\\
0 &0& 0&- r^4 \sin^2 \theta \left[1-2\left(\dfrac{M}{r}\right)^2\right]\\
    \end{pmatrix}{\cal E}_{2 m}Y_{2 m},
\end{align}
\end{widetext}
where ${\cal E}_{2m}Y_{2m}$ is given in Eq.~\eqref{eq:TidalStrength}. This expression is consistent with Eq.~(2) of Ref.~\cite{Cardoso:2021qqu} with Eqs.~(13) and~(14) therein.

\bibliography{Refs}


\onecolumngrid

\end{document}